\documentclass[12pt]{article}
\usepackage{amsfonts,epsfig,latexsym}
\usepackage{pstricks,pst-node}
\catcode`\@=11 \@addtoreset{equation}{section}
\def\theequation{\arabic{equation}}
\def\theequation{\thesection\arabic{equation}}

\newcommand{\dd}{\partial}

\newcommand{\tr}{{\rm tr \,}}

\newcommand{\be}{\begin{equation}}
\newcommand{\ee}{\end{equation}}
\newcommand{\ben}{\begin{displaymath}}
\newcommand{\een}{\end{displaymath}}
\newcommand{\ba}{\begin{eqnarray}}
\newcommand{\ea}{\end{eqnarray}}

\newcommand{\bean}{\begin{eqnarray*}}
\newcommand{\eean}{\end{eqnarray*}}

\def\@normalsize{\@setsize\normalsize{15pt}\xiipt\@xiipt
\abovedisplayskip 14pt plus3pt minus3pt%
\belowdisplayskip \abovedisplayskip
\abovedisplayshortskip  \z@ plus3pt%
\belowdisplayshortskip  7pt plus3.5pt minus0pt}
\def\small{\@setsize\small{13.6pt}\xipt\@xipt
\abovedisplayskip 13pt plus3pt minus3pt%
\belowdisplayskip \abovedisplayskip
\abovedisplayshortskip  \z@ plus3pt%
\belowdisplayshortskip  7pt plus3.5pt minus0pt
\def\@listi{\parsep 4.5pt plus 2pt minus 1pt
\itemsep \parsep \topsep 9pt plus 3pt minus 3pt}}
\def\underline#1{\relax\ifmmode\@@underline#1\else
$\@@underline{\hbox{#1}}$\relax\fi} \@twosidetrue \relax
\catcode`@=12
\catcode`\@=11

\def\section{\@startsection{section}{1}{\z@}{3.5ex plus 1ex minus
.2ex}{2.3ex plus .2ex}{\large\bf}}
\def\thesection{\arabic{section}.}

\def\ps@headings{\def\@oddfoot{}\def\@evenfoot{}
\def\@oddhead{\hbox{}\hfill
\makebox[.5\textwidth]{\raggedright\ignorespaces --\thepage{}--
\hfill }}
\def\@evenhead{\@oddhead}
\def\subsectionmark##1{\markboth{##1}{}} }
\ps@headings \catcode`\@=12 \relax

\def\figcap{\section*{Figure Captions\markboth
{FIGURECAPTIONS}{FIGURECAPTIONS}}\list {Fig.
\arabic{enumi}:\hfill}{\settowidth\labelwidth{Fig. 999:}
\leftmargin\labelwidth\advance\leftmargin\labelsep\usecounter{enumi}}}
\relax
\def\tablecap{\section*{Table Captions\markboth
{TABLECAPTIONS}{TABLECAPTIONS}}\list {Table
\arabic{enumi}:\hfill}{\settowidth\labelwidth{Table 999:}
\leftmargin\labelwidth
\advance\leftmargin\labelsep\usecounter{enumi}}}
 \relax
\def\reflist{\section*{References\markboth
{REFLIST}{REFLIST}}\list
{[\arabic{enumi}]\hfill}{\settowidth\labelwidth{[999]}
\leftmargin\labelwidth
\advance\leftmargin\labelsep\usecounter{enumi}}}
 \relax
\catcode`\@=11
\def\marginnote#1{}
\newcount\hour
\newcount\minute
\newtoks\amorpm
\hour=\time\divide\hour by60 \minute=\time{\multiply\hour by60
\global\advance\minute by- \hour}
\edef\standardtime{{\ifnum\hour<12 \global\amorpm={am}%
\else\global\amorpm={pm}\advance\hour by-12 \fi \ifnum\hour=0
\hour=12 \fi
\number\hour:\ifnum\minute<100\fi\number\minute\the\amorpm}}
\edef\militarytime{\number\hour:\ifnum\minute<100\fi\number\minute}
\def\draftlabel#1{{\@bsphack\if@filesw {\let\thepage\relax
\xdef\@gtempa{\write\@auxout{\string
\newlabel{#1}{{\@currentlabel}{\thepage}}}}}\@gtempa
\if@nobreak \ifvmode\nobreak\fi\fi\fi\@esphack}
\gdef\@eqnlabel{#1}}
\def\@eqnlabel{}
\def\@vacuum{}
\def\draftmarginnote#1{\marginpar{\raggedright\scriptsize\tt#1}}
\def\draft{\oddsidemargin -.5truein
\def\@oddfoot{\sl preliminary draft \hfil
\rm\thepage\hfil\sl\today\quad\militarytime}
\let\@evenfoot\@oddfoot \overfullrule 3pt
\let\label=\draftlabel
\let\marginnote=\draftmarginnote
\def\@eqnnum{(\theequation)\rlap{\kern\marginparsep\tt\@eqnlabel}%
\global\let\@eqnlabel\@vacuum}  }
\def\preprint{\twocolumn\sloppy\flushbottom\parindent 1em
\leftmargini 2em\leftmarginv .5em\leftmarginvi .5em \oddsidemargin
-.5in    \evensidemargin -.5in \columnsep 15mm \footheight 0pt
\textwidth 250mmin      \topmargin  -.4in \headheight 12pt
\topskip .4in \textheight 175mm \footskip 0pt
\def\@oddhead{\thepage\hfil\addtocounter{page}{1}\thepage}
\let\@evenhead\@oddhead \def\@oddfoot{} \def\@evenfoot{}  }
\def\titlepage{\@restonecolfalse\if@twocolumn\@restonecoltrue\onecolumn
\else \newpage \fi \thispagestyle{empty}\c@page\z@
\def\thefootnote{\fnsymbol{footnote}} }
\def\endtitlepage{\if@restonecol\twocolumn \else  \fi
\def\thefootnote{\arabic{footnote}}
\setcounter{footnote}{0}}  
\catcode`@=12 \relax

\def\ps@headings{\def\@oddfoot{}\def\@evenfoot{}
\def\@oddhead{\hbox{}\hfill
\makebox[.5\textwidth]{\raggedright\ignorespaces --\thepage{}--
\hfill }}
\def\@evenhead{\@oddhead}
\def\subsectionmark##1{\markboth{##1}{}} }
\ps@headings \relax\def\firstpage#1#2#3#4#5#6{
\begin{document}
\begin{titlepage}
\nopagebreak
\title{\begin{flushright}
\vspace*{-1.8in}
{\normalsize CPTH RR 057.0904}\\[-6mm]
{\normalsize LPT-ORSAY 04/82}\\[-6mm]
{\normalsize ROM2F-04/28}\\[-6mm]
{\normalsize hep-th/0410101}\\[-6mm]
\end{flushright}
\vfill {#3}}
\author{\large #4 \\[1.0cm] #5}
\maketitle \vskip -7mm \nopagebreak
\begin{abstract} {\noindent #6}
\end{abstract}
\vfill
\begin{flushleft}
\rule{16.1cm}{0.2mm}\\[-3mm]
$^{\dagger}${\small Unit{\'e} mixte du CNRS et de l'EP, UMR 7644.}\\[-3mm]
$^{\ddagger}${\small Unit{\'e} mixte du CNRS, UMR 8627.}\\
October \ 2004
\end{flushleft}
\thispagestyle{empty}
\end{titlepage}}
\def\simlt{\stackrel{<}{{}_\sim}}
\def\simgt{\stackrel{>}{{}_\sim}}
\newcommand{\dal}{\raisebox{0.085cm} {\fbox{\rule{0cm}{0.07cm}\,}}}
\newcommand{\dt}{\partial_{\langle T\rangle}}
\newcommand{\dtbar}{\partial_{\langle\overline{T}\rangle}}
\newcommand{\al}{\alpha^{\prime}}
\newcommand{\mst}{M_{\scriptscriptstyle \!S}}
\newcommand{\mpl}{M_{\scriptscriptstyle \!P}}
\newcommand{\dv}{\int{\rm d}^4x\sqrt{g}}
\newcommand{\lv}{\left\langle}
\newcommand{\rv}{\right\rangle}
\newcommand{\ph}{\varphi}
\newcommand{\abar}{\overline{a}}
\newcommand{\sbar}{\,\overline{\! S}}
\newcommand{\xbar}{\,\overline{\! X}}
\newcommand{\fbar}{\,\overline{\! F}}
\newcommand{\zbar}{\overline{z}}
\newcommand{\dbar}{\,\overline{\!\partial}}
\newcommand{\tbar}{\overline{T}}
\newcommand{\taubar}{\overline{\tau}}
\newcommand{\ubar}{\overline{U}}
\newcommand{\ybar}{\overline{Y}}
\newcommand{\phb}{\overline{\varphi}}
\newcommand{\cm}{Commun.\ Math.\ Phys.~}
\newcommand{\prl}{Phys.\ Rev.\ Lett.~}
\newcommand{\pr}{Phys.\ Rev.\ D~}
\newcommand{\pl}{Phys.\ Lett.\ B~}
\newcommand{\ibar}{\overline{\imath}}
\newcommand{\jbar}{\overline{\jmath}}
\newcommand{\np}{Nucl.\ Phys.\ B~}
\newcommand{\F}{{\cal F}}
\renewcommand{\L}{{\cal L}}
\newcommand{\A}{{\cal A}}
\newcommand{\e}{{\rm e}}
\newcommand{\dslash}{{\not\!\partial}}
\newcommand{\gsi}{\,\raisebox{-0.13cm}{$\stackrel{\textstyle >}{\textstyle\sim}$}\,}
\newcommand{\lsi}{\,\raisebox{-0.13cm}{$\stackrel{\textstyle <}{\textstyle\sim}$}\,}
\date{}
\firstpage{3118}{IC/95/34} {\large\bf On tadpoles and vacuum
redefinitions in String Theory} { E. Dudas$^{\,a,b}$, M.
Nicolosi$^{\,c}$, G. Pradisi$^{\,c}$ and A. Sagnotti$^{\,c}$}
{\\[-3mm]
\normalsize\sl $^a$ Centre de Physique
Th{\'e}orique$^\dagger$, Ecole Polytechnique, F-91128 Palaiseau, France\\[-2mm]
\normalsize\sl $^b$  LPT$^\ddagger$, B{\^a}t. 210, Univ. de
Paris-Sud, F-91405 Orsay,
France\\[-2mm]\normalsize\sl \normalsize\sl$^{c}$ Dipartimento di Fisica,
Univ. di Roma ``Tor
Vergata'' and INFN -- Sez. Roma II\\[-3mm]\normalsize\sl Via della Ricerca Scientifica 1, 00133 Roma,
Italy\\[-3mm]}
{Tadpoles accompany, in one form or another, all attempts to
realize supersymmetry breaking in String Theory, making the
present constructions at best incomplete. Whereas these tadpoles
are typically large, a closer look at the problem from a
perturbative viewpoint has the potential of illuminating at least
some of its qualitative features in String Theory. A possible
scheme to this effect was proposed long ago by Fischler and
Susskind, but incorporating background redefinitions in string
amplitudes in a systematic fashion has long proved very difficult.
In the first part of this paper, drawing from field theory
examples, we thus begin to explore what one can learn by working
perturbatively in a ``wrong'' vacuum. While unnatural in Field
Theory, this procedure presents evident advantages in String
Theory, whose definition in curved backgrounds is mostly beyond
reach at the present time. At the field theory level, we also
identify and characterize some special choices of vacua where
tadpole resummations terminate after a few contributions. In the
second part we present a notable example where vacuum
redefinitions can be dealt with to some extent at the full string
level, providing some evidence for a new link between IIB and 0B
orientifolds. We finally show that $NS$-$NS$ tadpoles do not
manifest themselves to lowest order in certain classes of string
constructions with broken supersymmetry and parallel branes,
including brane-antibrane pairs and brane supersymmetry breaking
models, that therefore have UV finite threshold corrections at one
loop.} \break

\section{Introduction}

In String Theory the breaking of supersymmetry is generally
accompanied by the emergence of $NS$-$NS$ tadpoles, one-point
functions for certain bosonic fields to go into the vacuum.
Whereas their $R-R$ counterparts signal inconsistencies of the
field equations or quantum anomalies \cite{pc}, these tadpoles are
commonly regarded as mere signals of modifications of the
background. Still, for a variety of conceptual and technical
reasons, they are the key obstacle to a satisfactory picture of
supersymmetry breaking, an essential step to establish a proper
connection with Particle Physics. Their presence introduces
infrared divergences in string amplitudes: while these have long
been associated to the need for background redefinitions
\cite{fs}, it has proved essentially impossible to deal with them
in a full-fledged string setting. For one matter, in a theory of
gravity these redefinitions affect the background space time, and
the limited technology presently available for quantizing strings
in curved spaces makes it very difficult to implement them in
practice.

This paper is devoted to exploring what can possibly be learnt if
one insists on working in a Minkowski background, that greatly
simplifies string amplitudes, even when tadpoles arise. This
choice may appear contradictory since, from the world-sheet
viewpoint, the emergence of tadpoles signals that the Minkowski
background becomes a ``wrong vacuum''. Indeed, loop and
perturbative expansions cease in this case to be equivalent, while
the leading infrared contributions need suitable resummations. In
addition, in String Theory $NS$-$NS$ tadpoles are typically large,
so that a perturbative approach is not fully justified. While we
are well aware of these difficulties, we believe that this
approach has the advantage of making a concrete string analysis
possible, if only of qualitative value in the general case, and
has the potential of providing good insights into the nature of
this crucial problem. A major motivation for us is that the
contributions to the vacuum energy from Riemann surfaces with
arbitrary numbers of boundaries, where $NS$-$NS$ tadpoles can
emerge already at the disk level, play a key role in orientifold
models \cite{orientifolds}. This is particularly evident for the
mechanism of brane supersymmetry breaking \cite{sugimoto,bsb},
where the simultaneous presence of branes and antibranes of
different types, required by the simultaneous presence of $O_+$
and $O_-$ planes, and possibly of additional brane-antibrane
systems \cite{sugimoto,bsb,aiq}, is generically accompanied by
$NS$-$NS$ tadpoles that first emerge at the disk and projective
disk level. Similar considerations apply to non-supersymmetric
intersecting brane models \cite{intersecting}\footnote{Or,
equivalently, models with internal magnetic fields.}, and the
three mechanisms mentioned above have a common feature: in all of
them supersymmetry is preserved, to lowest order, in the closed
sector, while it is broken in the open (brane) sector. However,
problems of this type are ubiquitous also in closed-string
constructions \cite{schschwstr} based on the Scherk-Schwarz
mechanism \cite{scherkschwarz}, where their emergence is only
postponed to the torus amplitude.

To give a flavor of the difficulties one faces, let us begin by
considering models where only a tadpole $\Delta^{(0)}$ for the
dilaton $\varphi$ is present. The resulting higher-genus
contributions to the vacuum energy are then plagued with infrared
(IR) divergences originating from dilaton propagators that go into
the vacuum at zero momentum, so that the leading (IR dominated)
contributions to the vacuum energy have the form
\ba \Lambda_0 &=& e^{-\varphi} \ \Delta^{(0)} +  \frac{1}{2}\,
\Delta^{(m)} \, \left( i\; {\cal D}_0^{mn}\right) \, \Delta^{(n)}
+ \frac{1}{2}\, e^{\varphi} \, \Delta^{(m)} \, \left( i\; {\cal
D}_0^{mn}\right) \, \Sigma^{np} \, \left( i\; {\cal
D}_0^{pq}\right) \, \Delta^{(q)} +
\cdots  \nonumber \\
&=&  e^{-\varphi} \ \Delta^{(0)} \ + \ \frac{1}{2}\, \Delta \, [\;
1-e^{\varphi} \left( i\; {\cal D}_0 \right) \Sigma \; ]^{-1}
\left( i\; {\cal D}_0 \right) \, \Delta \ + \cdots \ . \label{i1}
\ea
Eq.~(\ref{i1}) contains in general contributions from the dilaton
and from its massive Kaluza-Klein recurrences, implicit in its
second form, where they are taken to fill a vector $\Delta$ whose
first component is the dilaton tadpole $\Delta^{(0)}$. Moreover,
\be \langle m |{\cal D}_0 (p^2) |n \rangle \ \equiv \
 {\cal D}_0^{(mn)} (p^2) =
\delta^{mn} {\cal D}_0^{(mm)} (p^2)  \ee \label{i01} and \be
\langle m |\Sigma_0 (p^2) |n \rangle \ \equiv \ \Sigma^{(mn)}
(p^2) \label{i011} \ee
denote the sphere-level propagator of a dilaton recurrence of mass
$m$ and the matrix of two-point functions for dilaton recurrences
of masses $m$ and $n$ on the disk. They are both evaluated at zero
momentum in (\ref{i1}), where the first term is the disk
(one-boundary) contribution, the second is the cylinder
(two-boundary) contribution, the third is the genus 3/2
(three-boundary) contribution, and so on.  The resummation in the
last line of (\ref{i1}) is thus the string analogue of the more
familiar Dyson propagator resummation in Field Theory,
\be -i\; {\cal D}^{-1} (p^2) \ = \ -i\; {\cal D}_0^{-1} (p^2) -
e^{\varphi} \, \Sigma (p^2) \ , \label{i111} \ee
where in our conventions the self-energy $\Sigma (p^2)$ does not
include the string coupling in its definition. Even if the
individual terms in (\ref{i1}) are IR divergent, the resummed
expression is in principle perfectly well defined at zero
momentum, and yields
\be \Lambda_0 = e^{-\varphi}\Delta^{(0)} \, - \, \frac{1}{2}\,
e^{-\varphi} \, \Delta^{(0)} \, \Sigma^{-1 \ (00)} \, \Delta^{(0)}
\, +  \, \frac{1}{2} \sum_{m,n \neq 0} \! \Delta^{(m)}
\left([1-e^{\varphi} \left( i\; {\cal D}_0 \right) \Sigma ]^{-1}
\left( i\; {\cal D}_0 \right) \right)^{mn} \Delta^{(n)} \ .
\label{i02} \ee
In addition, the soft dilaton theorem implies that
\be e^\varphi \, \Sigma^{(00)} =
\frac{\partial}{\partial \varphi} \ \left(e^\varphi \,
\Delta^{(0)}\right) \ , \ee
so that the first two contributions cancel one another, up to a
relative factor of two. This is indeed a rather compact result,
but here we are describing for simplicity only a partial
resummation, that does not take into account higher-point
functions: a full resummation is in general far more complicated
to deal with, and therefore it is essential to identify possible
simplifications of the procedure.

A lesson we shall try to provide in this work, via a number of toy
examples  based on model field theories meant to shed light on
different features of the realistic string setting, is that
\emph{when a theory is expanded around a ``wrong'' vacuum, the
vacuum energy is typically driven to its value at a nearby
extremum (not necessarily a minimum)}, while the IR divergences
introduced by the tadpoles are simultaneously eliminated. In an
explicit example discussed in Section 2-b we also display some
wrong vacua in which higher-order tadpole insertions cancel both
in the field {\it v.e.v.} and in the vacuum energy, so that the
lowest corrections determine the full resummations. Of course,
subtle issues related to modular invariance or to its counterparts
in open-string diagrams are of crucial importance if this program
is to be properly implemented in String theory, and make the
present considerations somewhat incomplete. For this reason, we
plan to return to this key problem in a future publication
\cite{bdnps}, that will also include details of some string
computations whose results are displayed in subsection 3.2. The
special treatment reserved to the massless modes has nonetheless a
clear motivation: tadpoles act as external fields that in general
lift the massless modes, eliminating the corresponding infrared
divergences if suitable resummations are taken into account. On
the other hand, for massive modes such modifications are expected
to be less relevant, if suitably small. We present a number of
examples that are meant to illustrate this fact: small tadpoles
can at most deform slightly the massive spectrum, without any
sizable effect on the infrared behavior. The difficulty associated
with massless modes, however, is clearly spelled out in eq.
(\ref{i02}): resummations in a wrong vacuum, even within a
perturbative setting of small $g_s$, give rise to effects that are
typically large, of disk (tree) level, while the last term in
(\ref{i02}) due to massive modes is perturbatively small provided
the string coupling $e^\varphi$ satisfies the natural bound
$e^\varphi < m^2 / M_s^2$, where for the Kaluza-Klein case $m$
denotes the mass of the lowest recurrences and $M_s$ denotes the
string scale. The behavior of massless fields in simple models can
give a taste of similar difficulties that they introduce in String
Theory, and is also a familiar fact in Thermal Field Theory
\cite{tft}, where a proper treatment of IR divergences points
clearly to the distinct roles of two power-series expansions, in
coupling constants and in tadpoles. As a result, even models with
small couplings can well be out of control, and unfortunately this
is what happens in the most natural (and, in fact, in all known
perturbative) realizations of supersymmetry breaking in String
Theory. Some books treating the basics of these issues in Field
Theory are, for instance, \cite{books}.

Despite all these difficulties, at times string perturbation
theory can retain some meaning even in the presence of tadpoles.
For instance, in some cases one can identify subsets of the
physical observables that are insensitive to $NS$-$NS$ tadpoles.
There are indeed some physical quantities for which the IR effects
associated to the dilaton going into the vacuum are either absent
or are at least protected by perturbative vertices and/or by the
propagation of massive string modes. Two such examples are
threshold corrections to differences of gauge couplings for gauge
groups related by Wilson line breakings and scalar masses induced
by Wilson lines. For these quantities, the breakdown of
perturbation theory occurs at least at higher orders. There are
also models with ``small'' tadpoles. For instance, with suitable
fluxes \cite{fluxes} it is possible to concoct ``small'' tadpoles,
and one can then define a second perturbative expansion, organized
by the number of tadpole insertions, in addition to the
conventional expansion in powers of the string coupling
\cite{bdnps}.

In Section 2 we begin to gather some intuition on these matters
from toy models in Field Theory. After presenting the essentials
of the formalism in Subsection 2-a, in the following Subsection
2-b we discuss some simple explicit examples where the endpoint of
this process, that we call ``resummation flow'', is known, to
stress the type of subtleties associated with convergence domains
around inflection points of the scalar potential, where the
tadpole expansions break down. This type of considerations are in
principle of direct interest for String Theory, where the relevant
configuration space is very complicated, since, as we shall see, a
perturbative treatment runs the risk of terminating at a local
maximum. In Subsections 2-c and 2-d we perform explicit
resummations in field theories with tadpoles localized on
$D$-branes and $O$-planes of non-vanishing codimension, while in
Subsection 2-e we discuss, in a toy example, the inclusion of
gravity, that presents further subtleties related to the nature of
the graviton mass term. Section 3 contains some preliminary string
results. In Subsection 3-a we present an explicit example where
vacuum redefinitions can be performed explicitly at the string
level to some extent and provide some evidence that the correct
vacuum for a type-$II$ orientifold with local $R-R$ and $NS$-$NS$
tadpoles in compact internal dimensions is actually described by a
type-0 orientifold with a non-compact internal dimension, a
regular endpoint for a collapsing space time. In Subsection 3-b we
begin to investigate the sensitivity of quantum corrections to
$NS$-$NS$ tadpoles. We show, in particular, that in
non-supersymmetric models with parallel branes, that include brane
supersymmetry breaking models and models with brane-antibrane
pairs, quantum corrections are UV finite at one-loop, and are
therefore insensitive to the tadpoles. Finally, the conclusions
summarize our present understanding of the problem, in particular
for what concerns the behavior at higher orders.

\section{Quantum Field Theory in a wrong vacuum}

\subsection{``Wrong vacua'' and the effective action}

The standard formulation of Quantum Field Theory refers implicitly
to a choice of ``vacuum'', instrumental for defining the
perturbative expansion, whose key ingredient is the generating
functional of connected Green functions $W[J]$. Let us refer for
definiteness to a scalar field theory, for which
\be e^{\frac{i}{\hbar}W[J]} \ = \ \int\,[ D\phi ] \,
e^{\,\frac{i}{\hbar}\,(S(\phi)+  \int d^{\cal D} x \, J \phi )} \
, \ee
where
\be S(\phi) \ = \ \int d^{\cal D} x \left(\; - \;
\frac{1}{2}\,\partial_\mu\phi\,\partial^\mu\phi-V(\phi)\right) \ ,
\ee
written here symbolically for a collection of scalar fields $\phi$
in ${\cal D}$ dimensions. Whereas the conventional saddle-point
technique rests on a shift
\be \phi \ = \ \varphi \ + \ \phi_0 \ee
about an extremum of the full action with the external source,
here we are actually interested in expanding around a ``wrong''
vacuum $\phi_0$, defined by
\be \left. \frac{\delta S}{\delta \phi}\right|_{\phi=\phi_0} \ = \
- \; (J + \Delta ) \ , \label{ft9} \ee
where, for simplicity, in this Subsection we shall let $\Delta$ be
a constant, field-independent quantity, to be regarded as the
classical manifestation of a tadpole. In the following
Subsections, however, we shall also discuss examples where
$\Delta$ depends on $\phi_0$.

The shifted action then expands according to
\be S(\varphi + \phi_0) \ = \ S(\phi_0) \ - \  (J + \Delta) \,
\varphi \ + \ \frac{1}{2} \, \varphi \, \left( i\, {\cal D}^{-1}
\right)_{\phi_0} \, \varphi \ + \ S_I(\phi_0,\varphi) \ ,
\label{ft1} \ee where \be \left( i\, {\cal D}^{-1}\right)_{\phi_0}
\ = \ \left. {\delta^2 \ S \over \delta \phi^2}\right|_{\phi_0} \
, \label{ft2} \ee
while $S_I(\phi_0,\varphi)$ denotes the interacting part, that in
general begins with cubic terms. After the shift, the generating
functional
\be e^{\frac{i}{\hbar}W[J]}=e^{\frac{i}{\hbar}[S(\phi_0)+ J\phi_0
]}\int [ D\varphi ] \, e^{\frac{i}{\hbar}[\frac{1}{2} \varphi \; i
\mathcal{D}^{-1}(\phi_0) \; \varphi - \Delta \varphi +
S_I(\varphi,\phi_0)]}  \label{ft3} \ee
can be put in the form
\be W [J] = S(\phi_0 ) +  \phi_0 J  +  \frac{i\hbar}{2}
\tr\ln\left(i\left. \mathcal{D}^{-1} \right|_{\phi_0}\right) + W_2
[J] + {i \over 2} \Delta \mathcal{D} \Delta \ , \label{ft4} \ee
where \be e^{\frac{i}{\hbar}W_2[J]}=\frac{\int [ D\varphi ] \,
e^{\frac{i}{\hbar}[\frac{1}{2}\varphi \; i \mathcal{D}^{-1}
(\phi_0)\; \varphi - \varphi \Delta + S_I(\varphi,\phi_0)]}} {\int
[ D\varphi ] \, e^{\frac{i}{\hbar} ({1 \over 2} \varphi \; i
\mathcal{D}^{-1} \; \varphi - \varphi \Delta )}} \ .
 \label{ft5} \ee

In the standard approach, $W_2$ is computed perturbatively
\cite{cw}, expanding $\exp(\frac{iS_I}{\hbar})$ in a power series,
and contributes starting from two loops. On the contrary, if
classical tadpoles are present it also gives tree-level
contributions to the vacuum energy, but these are at least ${\cal
O}(\Delta^3)$. Defining as usual the effective action as
\be \Gamma(\bar\phi) \ = \ W[J] \ - \ J\bar\phi \ , \label{ft6}
\ee
the classical field is
\be \bar\phi=\frac{\delta W}{\delta J}= \phi_0 + \frac{\delta
\phi_0}{\delta J}\,\bigg(\; - \; \Delta+\frac{1}{2}
\Delta^2\frac{\delta}{\delta \phi_0} i \mathcal{D}(\phi_0)+
\frac{i \hbar}{2}\tr\frac{\delta}{\delta \phi_0}
\ln(i\mathcal{D}^{-1}({\bar \phi_0})) \bigg) + \cdots \ .
\label{ft12} \ee
Notice that $\bar\phi$ is no longer a small quantum correction to
the original ``wrong'' vacuum configuration $\phi_0$, and indeed
the second and third terms on the \emph{r.h.s.} of (\ref{ft12}) do
not carry any $\hbar$ factors. Considering only tree-level terms
and working to second order in the tadpole, one can solve for
$\phi_0$ in terms of $\bar\phi$, obtaining
\be \phi_0\ = \ \bar\phi \ + \ i \mathcal{D}(\bar\phi)\Delta \ + \
\frac{1}{2}\,i \mathcal{D}(\bar\phi)
\,\frac{\delta}{\delta\bar\phi}i\mathcal{D}(\bar\phi)\,\Delta^2 \
+ \ {\cal O}(\Delta^3) \ ,\ee
and substituting in the expression for $\Gamma[\bar\phi]$ then
gives
\be \Gamma[\bar\phi] \ = \ S(\bar\phi)+\left( \left. {\delta \ S
\over \delta \phi}\right|_{\bar{\phi}}+J\right)\,
\left[i\mathcal{D}(\bar\phi)\,\Delta+
\frac{1}{2}\,i\mathcal{D}(\bar\phi)\,\frac{\delta}{\delta\bar\phi}\,
i\mathcal{D}(\bar\phi)\,\Delta^2\right]+{\cal O}(\Delta^3) \ .
\label{gammawrong} \ee
One can now relate $J$ to $\,\bar\phi\,$ using eq.~(\ref{ft9}),
and the result is
\be J  = -\; \left. {\delta \ S \over \delta
\phi}\right|_{\bar{\phi}}\, -i\; \mathcal{D}^{-1}\,
(\bar\phi)(i\mathcal{D}(\bar\phi)\,\Delta) -\Delta+{\cal
O}(\Delta^2)=-\left. {\delta \ S \over \delta
\phi}\right|_{\bar{\phi}}+{\cal O}(\Delta^2) \ . \ee
Making use of this expression in (\ref{gammawrong}), all explicit
corrections depending on $\Delta$ cancel, and the tree-level
effective action reduces to the classical action:
\be \Gamma[\bar\phi] \ = \ S(\bar\phi)\ + \ {\cal O}(\Delta^3) \ .
\ee \label{wrongright}
This is precisely what one would have obtained expanding around an
extremum of the theory, but we would like to stress that this
result is here recovered expanding around a ``wrong'' vacuum. The
terms ${\cal O}(\Delta^3)$, if properly taken into account, would
also cancel against other tree-level contributions originating
from $W_2[J]$, so that the recovery of the classical vacuum energy
would appear to be exact. We shall return to this important issue
in the next Section.

\begin{figure}[h]
\begin{center}
  \resizebox{14cm}{!}{\psfig{figure=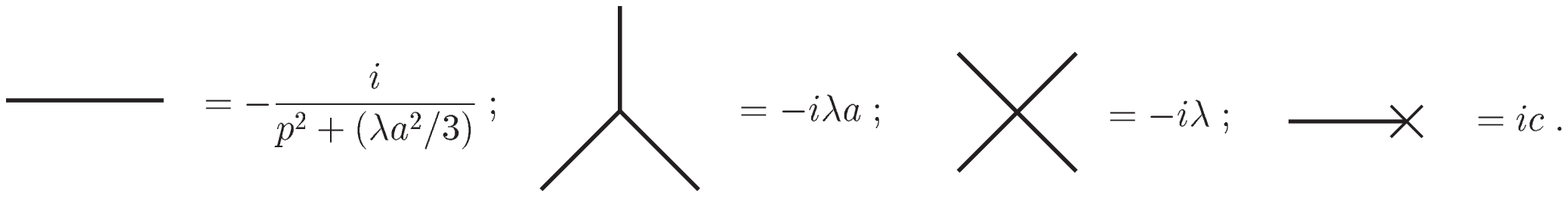,width=14cm}}
\caption{Feynman rules at $\phi_0=a$} \label{fig3}
\end{center}
\end{figure}

\begin{figure}[h]
\begin{center}
  \resizebox{14cm}{!}{\psfig{figure=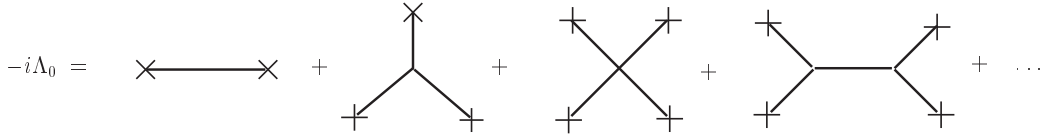,width=14cm}}
\caption{Vacuum energy} \label{fig4}
\end{center}
\end{figure}

Let us take a closer look at the case of small tadpoles, that is
naturally amenable to a perturbative treatment. This can
illustrate a further important subtlety: the diagrams that
contribute to this series \emph{are not all 1PI}, and thus by the
usual rules \emph{should not all} contribute to $\Gamma$. For
instance, let us consider the Lagrangian
\be {\cal L} \ =\  - \, \frac{1}{2} \, (\dd_\mu\phi)^2\ -\
\frac{\lambda}{4!}(\phi^2-a^2)^2 \ + \ c\; \phi \ ,
\label{modelaction} \ee
with a Mexican-hat section potential and a driving ``magnetic
field'' represented by the tadpole $c$. The issue is to single out
the terms produced in the gaussian expansion of the path integral
of eq. (\ref{modelaction}) once the integration variable is
shifted about the ``wrong'' vacuum $\phi_0=a$, writing $\phi = a+
\chi$, so that
\be e^{\frac{i}{\hbar} \, W[J]} \ = \int [D\phi] \
e^{\frac{i}{\hbar} \int d^{\cal D} x \left( \; - \; \frac{1}{2}
\partial_\mu \chi
\partial^\mu \chi \, - \, \frac{\lambda a^2}{6} \chi^2  \, - \,
\frac{\lambda a}{3!} \chi^3 \, - \, \frac{\lambda}{4!} \chi^4 \, +
\, c(a+ \chi) \, + \, J (a + \chi) \right)} \ . \ee
Notice that, once $\chi$ is rescaled to $\hbar^{1/2} \chi$ in
order to remove all powers of $\hbar$ from the gaussian term, in
addition to the usual positive powers of $\hbar$ associated to
cubic and higher terms, a negative power $\hbar^{-1/2}$
accompanies the tadpole term in the resulting Lagrangian. As a
result, the final ${\cal O}(1/\hbar)$ contribution that
characterizes the classical vacuum energy results from infinitely
many diagrams built with the Feynman rules summarized in fig.
\ref{fig3}. The first two non-trivial contributions originate from
the three-point vertex terminating on three tadpoles, from the
four-point vertex terminating on four tadpoles and from the
exchange diagram of fig. \ref{fig4}.

As anticipated, tadpoles affect substantially the character of the
diagrams contributing to $\Gamma$, and in particular to the vacuum
energy, that we shall denote by $\Lambda_0$. Beginning from the
latter, let us note that, in the presence of a tadpole coupling
$c$,
\be e^{- \frac{i}{\hbar} \Lambda_0\, {\cal V}} \ = \ \int [D\phi]
\ e^{\frac{i}{\hbar} \left( S[\phi] \, + \, \int d^{\cal D} x \; c
\phi \right)} \ , \ee
where ${\cal V}$ denotes the volume of space time. Hence, the
vacuum energy is actually determined by a power series in $c$
whose coefficients are \emph{connected}, rather than 1PI,
amplitudes, since they are Green functions of W computed for a
classical value of the current determined by $c$:
\be - \, \frac{i}{\hbar} \, \Lambda_0 \, {\cal V} \  = \ \sum_n \
\frac{(i \; c )^n}{n! \; \hbar^n} \ W^{(n)}[\{p_j = 0\}] \ . \ee
A similar argument applies to the higher Green functions
of $\Gamma$: the standard Legendre transform becomes effectively
in this case
\be W[J+c] \ = \ \Gamma[\bar{\phi}] \ + \ J
\bar{\phi} \ , \label{shiftleg} \ee
since the presence of a tadpole shifts the argument of W. However,
the \emph{l.h.s.} of (\ref{shiftleg}) contains an infinite series
of conventional connected Green functions, and after the Legendre
transform only those portions that do not depend on the tadpole
$c$ are turned into 1PI amplitudes. The end conclusion is indeed
that the contributions to $\Gamma$ that depend on the tadpoles
involve arbitrary numbers of connected, but also non 1PI,
diagrams.

The vacuum energy is a relatively simple and most important
quantity that one can deal with from this viewpoint, and its
explicit study will help to clarify the meaning of
eq.~(\ref{gammawrong}). Using eqs. (\ref{ft4}) and (\ref{ft6}) one
can indeed conclude that, at the classical level,
\be \Lambda_0 \ = \ - \, \frac{W(J=0)}{\cal V} \ = \ - \, V
(\phi_0) \ - \ \frac{i}{2} \ \Delta \, \left. {\cal
D}\right|_{p^2=0} \Delta + {\cal O} (\Delta^3) \ , \label{lambda1}
\ee
an equation that we shall try to illustrate via a number of
examples in this paper. The net result of this Subsection is that
resummations around a wrong vacuum lead nonetheless to extrema of
the effective action. However, it should be clear from the
previous derivation that the scalar propagator must be
nonsingular, or equivalently the potential must not have an
inflection point at $\phi_0$, in order that the perturbative
corrections about the original wrong vacuum be under control. A
related question is whether the resummation flow converges
generically towards minima (local or global) or can end up in a
maximum. As we shall see in detail shortly, the end point is
generally an extremum and not necessarily a local minimum.

\subsection{The end point of the resummation flow}

The purpose of this Section is to investigate, for some explicit
forms of the scalar potential $V (\phi)$ and for arbitrary initial
values of the scalar field $\phi_0$, the end point reached by the
system after classical tadpole resummations are performed. The
answer, that will be justified in a number of examples, is as
follows: {\it starting from a wrong vacuum $\phi_0$, the system
typically reaches a nearby extremum (be it a minimum or a maximum)
of the potential not separated from it by any inflection.} While
this is the generic behavior, we shall also run across a notable
exception to this simple rule: there exist some peculiar ``large''
flows, corresponding to special values of $\phi_0$, that can
actually reach an extremum by going past an inflection, and in
fact even crossing a barrier, but are nonetheless captured by the
low orders of the perturbative expansion!

An exponential potential is an interesting example that is free of
such inflection points, and is also of direct interest for
supersymmetry breaking in String Theory. Let us therefore begin by
considering a scalar field with the Lagrangian
\be {\cal L} \ = \ - \, \frac{1}{2} \, \partial^\mu \phi \,
\partial_\mu \phi \ - \ \alpha \ e^{b\phi} \ , \ee
where for definiteness the two coefficients $\alpha$ and $b$ are
both taken to be positive. The actual minimum is reached as $\phi
\to - \infty$, where the classical vacuum energy vanishes.

\begin{figure}[h]
\begin{center}
\resizebox{14cm}{!}{\psfig{figure=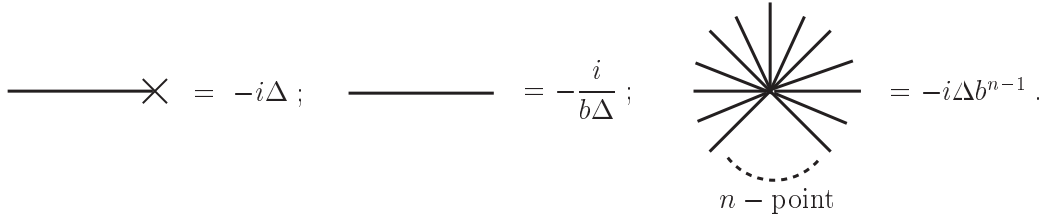,width=14cm}}
\caption{Feynman rules for the exponential potential} \label{fig5}
\end{center}
\end{figure}

In order to recover this result from a perturbative expansion
around a generic ``wrong'' vacuum $\phi_0$, let us shift the
field, writing $\phi=\phi_0+\chi$. The Feynman rules can then be
extracted from
\be {\cal L}_{eff} = - \, \frac{1}{2}\
\partial^\mu \chi \, \partial_\mu \chi \ - \ \frac{\Delta}{b}\
e^{b\chi} \ , \ee
where $\Delta$, the one-point function in the ``wrong'' vacuum, is
defined by
\be - \ \left . \frac{\delta {\cal
L}_{eff}}{\delta\phi}\right|_{\phi_0} \ = \ \Delta \ = \ \alpha \
b \ e^{b \phi_0} \ , \ee
and the first few contributions to the classical vacuum energy are
as in fig. \ref{fig5}.

It is fairly simple to compute the first few diagrams. For
instance, the two-tadpole correction to $- i V(\phi_0)$ is
\be \ \frac{1}{2} \ \frac{-i}{b\Delta} \ (-i\Delta)^2 =
-\frac{1}{2}\left(\frac{-i\Delta}{b}\right) \ , \ee while the
three-tadpole correction, still determined by a single diagram, is
\be \frac{1}{3!} \ (-i \Delta \ b^2) \ (-i \Delta)^3
\left(\frac{-i}{b \Delta}\right)^3 \ = \ - \
\frac{1}{6}\left(\frac{-i\Delta}{b}\right) \ . \ee
On the other hand, the quartic contribution is determined by two
distinct diagrams, and equals
\be \frac{1}{4!} \ (-i \Delta \ b^3) \ (-i \Delta )^4
\left(\frac{-i}{b \Delta}\right)^4 + \frac{1}{8} \ (-i \Delta \
b^2)^2 \ (-i \Delta)^4 \left(\frac{-i}{b \Delta}\right)^5 \ = \
 -\ \frac{1}{12}\left(\frac{-i\Delta}{b}\right) \ , \ee
while the quintic contribution originates from three diagrams.
Putting it all together, one obtains
\be \Lambda_0 \ = \ \frac{\Delta}{b} \
\left(1-\frac{1}{2}-\frac{1}{6}-\frac{1}{12}-\frac{1}{20}+\ldots
\right) \ = \ \frac{\Delta}{b}\left[1- \sum_{n=1}^{\infty}
\frac{1}{n(n+1)} \right] \ . \label{expwrong} \ee
The resulting pattern is clearly identifiable, and suggests in an
obvious fashion the series in (\ref{expwrong}). Notice that,
despite the absence of a small expansion parameter, in this
example the series in (\ref{expwrong}) actually \emph{converges}
to $1$, so that the correct vanishing value for the classical
vacuum energy can be exactly recovered from an arbitrary wrong
vacuum $\phi_0$.

\begin{figure}[h]
\begin{center}
\resizebox{5cm}{!}{\psfig{figure=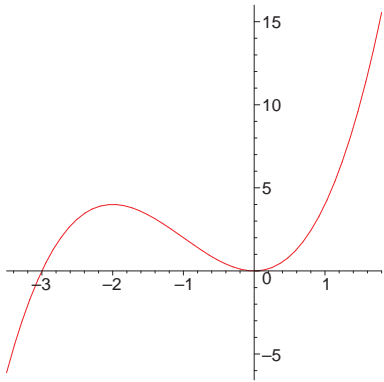,width=5cm}} \caption{A
cubic potential} \label{cubic}
\end{center}
\end{figure}

We can now turn to a more intricate example and consider the model
(fig. \ref{cubic})
\be {\cal L} \ = \ - \, \frac{1}{2} (\dd_\mu\phi)^2\ -\  \frac{m^2
\phi^2}{2} \ - \ \frac{\lambda \phi^3}{6} \ , \label{ft33} \ee
the simplest setting where one can investigate the role of an
inflection. Strictly speaking this example is pathological, since
its Hamiltonian is unbounded from below, but for our purpose of
gaining some intuition on classical resummations it is nonetheless
instructive. The two extrema of the scalar potential $v_{1,2}$ and
the inflection point $v_I$ are
\be v_1 = 0 \ , \quad v_2 = - \frac{2 m^2}{\lambda} \ , \quad v_I
= - \frac{ m^2}{\lambda} \ . \label{ft14} \ee
Starting from an arbitrary initial value $\phi_0$, let us
investigate the convergence of the resummation series and the
resulting resummed value $\langle \phi \rangle$. A close look at
the diagrammatic expansion indicates that
\ba && \langle \phi \rangle \ = \ \phi_0 + \ \frac{V'}{V^{''}} \
\sum_{n=0}^{\infty} c_n \left[ \frac {\lambda V'}{(V^{''})^2}
\right]^n \ \equiv \ \phi_0 + \frac{V'}{V^{''}} \ f (x) \ ,
\label{ft15}\ea
where the actual expansion variable is
\be x \ = \ \frac {\lambda V'}{(V^{''})^2} \ , \label{xvar} \ee
to be contrasted with the naive dimensionless expansion variable
\be z = \frac{\lambda \phi_0}{m^2} \ . \ee
According to eq.~(\ref{xvar}), their relation is
\be x = \frac{z(z+2)}{2(z+1)^2} \ , \ee
whose inverse is
\be z = -1 \ \mp \ \frac{1}{\sqrt{1 - 2 x }} \ , \label{zdouble}
\ee
where the upper sign corresponds to the region $(L)$ to the left
of the inflection, while the lower sign corresponds to the region
$(R)$ to the right of the inflection. In other words: $\phi_0$,
and thus the naive variable $z$ of the problem, is actually a
double-valued function of $x$, while the actual range covered by
$x$ terminates at the inflection.

A careful evaluation of the symmetry factors of various diagrams
with variable numbers of tadpole insertions shows that
\be
 f (x) \ = \ - \, \frac{\sqrt{\pi}}{2} \sum_{n=0}^{\infty} \frac{(-1)^n
   \ 2^{n+1}}{(n+1)!\ \Gamma {(1/2-n)}} \ x^n\ ,
\label{ft166} \ee
a series that for $|x|<1/2$ converges to
\be
 f(x)  \ = \ -\, \frac{1}{x} \ + \
 \frac{\sqrt{1-2x}}{x} \ . \label{ft16}
\ee
The relation between $z$ and $x$ implies that both $\phi_0$ and
$V^{''}$ have two different expressions in terms of $x$ on the two
sides of the inflection,
\be V^{''} = \pm \frac{m^2}{\sqrt{1-2x}} \quad , \quad \phi_0 =
-\frac{m^2}{\lambda} \mp \frac{m^2}{\lambda} \frac{1}{\sqrt{1-2x}}
\ , \label{ft17} \ee
where the upper signs apply to the region $(L)$ that lies to the
left of the inflection, while the lower signs apply to the region
$(R)$ that lies to the right of the inflection. Combining
(\ref{ft16}) and (\ref{ft17}) finally yields the announced result:
\ba && \langle \phi \rangle \rightarrow 0 \quad {\rm in \
the \ (R) \ region: } \quad - \frac{m^2}{\lambda}  < \phi_0 < \infty \ , \nonumber \\
&& \langle \phi \rangle \rightarrow - \frac{2m^2}{\lambda} \quad
 {\rm for \ in \ (L) \ region:} \quad - \infty < \phi_0 < - \frac{m^2}{\lambda} \
 . \label{ft18}
\ea
The resummation clearly breaks down near the inflection point
$v_I$. In the present case, the series in (\ref{ft16}) converges
for $|x| < 1/2$, and this translates into the condition
\be \phi_0 \in \left(- \infty,-\frac{m^2}{\lambda} \left(1 +
\frac{1}{\sqrt{2}}\right)\right) \ \bigcup \
\left(-\frac{m^2}{\lambda} \left(1 -
\frac{1}{\sqrt{2}}\right),\infty\right) \
 . \label{ft19} \ee

A symmetric interval around the inflection point thus lies outside
this region, while in the asymptotic regions $\phi_0 \rightarrow
\pm \infty$ the parameter $x$ tends to $1/2$, a limiting value for
the convergence of the series (\ref{ft166}).

The vacuum energy $\Lambda_0$ is another key quantity for this
problem. Starting as before from an arbitrary initial value
$\phi_0$, standard diagrammatic methods indicate that
\be \Lambda_0 = \ V(\phi_0) + \ \frac{(V')^2}{V^{''}} \
\sum_{n=0}^{\infty} d_n \left[ \frac {\lambda V'}{(V^{''})^2}
\right]^n \ \equiv \ V(\phi_0) + \frac{(V')^2}{V^{''}} \ h (x) \ .
\label{ft20} \ee
A careful evaluation of the symmetry factors of various diagrams
with arbitrary numbers of tadpole insertions then shows that
\be
 h (x) \ = \ - \, \frac{\sqrt{\pi}}{4} \sum_{n=0}^{\infty} \frac{(-1)^n
   \ 2^{n+2}}{(n+2)!\ \Gamma {(1/2-n)}} \ x^n  \ , \label{ft21} \ee
that for $|x| <1/2$ converges to
\be h(x) \ = \ \frac{1}{3x^2} \left[1-3x - (1-2x)^{3/2} \right] \
. \ee

The two different relations between $z$ and $x$ in (\ref{zdouble})
that apply to the two regions $(L)$ and $(R)$ finally yield:
\ba && \Lambda_0 \rightarrow 0 \ , \quad {\rm in \ the \ (R) \
  region } \ - \frac{m^2}{\lambda}  < \phi_0 < \infty \ , \nonumber \\
&& \Lambda_0 \rightarrow \frac{2m^6}{3 \lambda^2} \ , \quad
 {\rm in \ the \ (L) \ region} \   -\infty < \phi_0 < - \frac{m^2}{\lambda} \
 . \label{ft22}
\ea
To reiterate, we have seen how in this model the resummations
approach nearby extrema (local minima or maxima) not separated
from the initial value $\phi_0$ by any inflection.

\begin{figure}[h]
\begin{center}
\resizebox{5cm}{!}{\psfig{figure=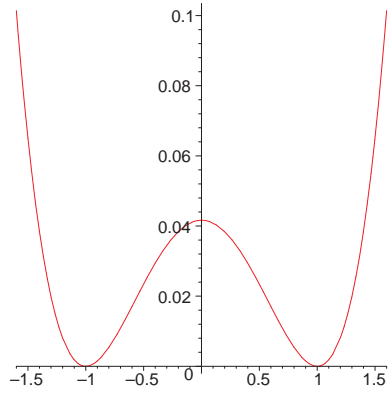,width=5cm}}
\caption{A quartic potential} \label{quartic}
\end{center}
\end{figure}

A physically more interesting example is provided by a real scalar
field described by the Lagrangian (fig. \ref{quartic})
\be {\cal L} \ = \ - \, \frac{1}{2} (\dd_\mu\phi)^2 \ - \
\frac{\lambda}{4!}\; (\phi^2-a^2)^2 \ , \label{phitothefour} \ee
whose potential has three extrema, two of which are a pair of
degenerate minima, $v_1= -a$ and $v_3=+a$, separated by a
potential wall, while the third is a local maximum at the origin.
Two inflection points are now present,
\be v^{(1)}_I = - \frac{a}{\sqrt{3}} \ , \qquad v^{(2)}_I =
\frac{a}{\sqrt{3}} \ , \ee
that also form a symmetric pair with respect to the vertical axis.

Starting from an arbitrary initial value $\phi_0$, one can again
in principle sum all the diagrams, and a closer look shows that
there are a pair of natural variables, $y_1$ and $y_2$, defined as
\be y_1 \ = \ \phi_0 \ \frac {\lambda V'}{(V^{''})^2} \qquad {\rm
and} \qquad y_2 \ =\ \frac {\lambda {V'}^2}{(V^{''})^3} \ ,
\label{ft23} \ee
that reflect the presence of the cubic and quartic vertices and
depend on the square of $\phi_0$. On the other hand, the naive
dimensionless variable to discuss the resummation flow is in this
case
\be z=\phi_0^2 / {a}^2 \ , \ee
and eqs.~(\ref{ft23}) imply that
\ba y_1 &=& 6\, \frac{z(z-1)}{(3 z -1 )^2} \ , \\
y_2 &=& 6\, \frac{z(z-1)^2}{(3 z -1 )^3} \ . \ea

The diagrammatic expansions of $\langle \phi \rangle$ and
$\Lambda_0$ are now more complicated than in the $\phi^{3}$
example. A careful evaluation of the symmetry factors of various
diagrams, however, uncovers an interesting pattern, since
\ba \langle \phi \rangle &=& \phi_0 + \frac{V'}{V^{''}} \left[-1
-\frac{1}{2}\left(y_1-\frac{y_2}{3}\right)-
\frac{1}{2}\left(y_1-\frac{y_2}{3}\right)
\left(y_1-\frac{y_2}{2}\right) \right. \nonumber
\\&-& \left. \frac{5}{8}\left(y_1-\frac{y_2}{3}\right) \left(y_1 - \frac{2 y_2}{3}\right)\left(y_1 -
\frac{2 y_2}{5}\right) \right. \nonumber \\
&-& \left. \frac{7}{8} \left(y_1 - \frac{y_2}{3}\right) \left(
y_1^3 - \frac{5}{3} y_1^2 y_2 + \frac{55}{63} y_1 y_2^2 -
\frac{55}{378} y_2^3 \right) \right. \nonumber \\
&-& \frac{21}{16}\ \left.
\left(y_1-\frac{y_2}{3}\right)\left(y_1^4-\frac{16}{7} y_1^3
y_2+\frac{13}{7} y_1^2 y_2^2 -
\frac{52}{81} y_1 y_2^3+\frac{13}{162} y_2^4\right)  + \cdots \right] \ , \nonumber \\
\Lambda_0 &=& V(\phi_0) + \frac{(V')^2}{V^{''}} \left[-\frac{1}{2}
-\frac{1}{6}\left(y_1-\frac{y_2}{4}\right)-
\frac{1}{8}\left(y_1-\frac{y_2}{3}\right)^2-
\frac{1}{8}\left(y_1-\frac{y_2}{3}\right)^2
\left(y_1- \frac{1}{2} y_2\right) \right. \nonumber \\
&-& \frac{7}{48} \left. \left(y_1-\frac{y_2}{3}\right)^2 \, \left(
y_1^2 - \frac{22}{21}
y_1 y_2 + \frac{11}{42} y_2^2 \right) \right. \nonumber \\
&-& \left. \frac{3}{16}\left(y_1-\frac{y_2}{3}\right)^2
\left(y_1^3-\frac{13}{8} y_1^2 y_2+ \frac{91}{108} y_1
y_2^2-\frac{91}{648} y_2^3\right) + \cdots \right] \ ,
\label{ft24} \ea
where all linear and higher-order corrections in $\langle \phi
\rangle$ and all quadratic and higher order corrections in
$\Lambda_0$ apparently disappear at the special point $y_1=y_2/3$.
Notice that this condition identifies the three extrema
$\phi_0=\pm a$ and $\phi_0=0$, but also, rather surprisingly, the
two additional points $\phi_0 = \pm \frac{a}{2}$. In all these
cases the series expansions for $\langle \phi \rangle$ and
$\Lambda_0$ apparently end after a few terms.

If one starts from a wrong vacuum {\it sufficiently close} to one
of the extrema, one can convince oneself that, in analogy with the
previous example \footnote{The simple pattern in eq. (2.52)
applies to regions I and III. Region II, however, has a richer
structure and includes three distinct subregions. For $-a/\sqrt{5}
< \phi_0 < a/\sqrt{5}$, the resummation flow does converge to
$v=0$, but the points $\phi_0 = \pm a/\sqrt{5}$ are very peculiar.
Indeed, starting from $\phi_0 = a/\sqrt{5}$, the first iteration
of the tangent method yields $\phi^{(1)} = - a/\sqrt{5}$, while
the second iteration gives again $\phi^{(2)} =  a/\sqrt{5}$, so
that the resummation flow oscillates between these two points
without converging to any extremum. For $a/\sqrt{5} < \phi_0 < a
y$, where $y $ is defined by the algebraic equation $2 \sqrt{3}
y^3 + 3 y^2 -1=0$, the resummation flow approaches the correct
miminum $<\phi> = a$. The point $\phi_0 = a y$ is defined by the
condition that the first iteration lead precisely to the
inflection point $\phi^{(1)} = - a/\sqrt{3}$. Finally, in the
third subregion $a y < \phi_0 < a/\sqrt{3}$, that contains in
particular the non-renormalisation point $\phi_0 = a/2$, the
resummation flow crosses the barrier and converges to $<\phi> =
-a$. The last two regions have of course mirror counterparts
obtained for $\phi_0 \rightarrow -\phi_0 $. These considerations
also apply in the presence of a small magnetic field. We are very
grateful to W. Mueck for calling these subtleties of Region II to
our attention.},
\ba && \langle \phi \rangle \rightarrow -a \quad , \quad {\rm for
 \
  region \ 1:} \quad -\infty < \phi_0 <  v^{(1)}_I\ , \nonumber \\
&& \langle \phi \rangle \rightarrow 0 \quad , \quad
 {\rm for \ region \ 2:} \quad v^{(1)}_I  < \phi_0 < v^{(2)}_I \
 , \nonumber \\
&& \langle \phi \rangle \rightarrow +a \quad , \quad
 {\rm for \ region \ 3:} \quad v^{(2)}_I   < \phi_0 < \infty \  ,
\label{regions} \ea
but we have not arrived at a single natural expansion parameter
for this problem, an analog of the variable $x$ of the cubic
potential. In addition, while these perturbative flows follow the
pattern of the previous example, since they are separated by
inflections that act like barriers, a puzzling and amusing result
concerns the special initial points
\be
 \phi_0 = \pm \frac{a}{2} \ . \label{nonre}
\ee
In this case $y_2 = 3 y_1$ and, as we have seen, apparently all
but the first few terms in $\langle \phi \rangle$ and all but the
first few terms in $\Lambda_0$ vanish. The non-vanishing terms in
eq.~(\ref{ft24}) show explicitly that the endpoints of these
resummation flows correspond to $\langle \phi \rangle = \pm a$ for
$\phi_0 = \mp \frac{a}{2}$, and that $ \Lambda_0 = 0$, so that
these two flows apparently ``cross'' the potential barrier and
pass beyond an inflection. One might be tempted to dismiss this
phenomenon, since after all this is a case with large tadpoles
(and large values of $y_1$ and $y_2$), that is reasonably outside
the region of validity of perturbation theory and hence of the
strict range of applicability of eq. (\ref{ft24}). Still, toward
the end of this Subsection we shall encounter a similar
phenomenon, clearly within a perturbative setting, where the
resummation will unquestionably collapse to a few terms to land at
an extremum, and therefore it is worthwhile to pause and devote to
this issue some further thought.

Interestingly, the tadpole resummations that we are discussing
have a simple interpretation in terms of Newton's method of
tangents, a very effective iterative procedure to derive the roots
of non-linear algebraic equations. It can be simply adapted to our
case, considering the function $V'(\phi)$, whose zeroes are the
extrema of the scalar potential. The method begins with guess, a
``wrong vacuum'' $\phi_0$, and proceeds via a sequence of
iterations determined by the zeros of the sequence of straight
lines
\be y - V' (\phi^{(n)}) \ = \ V^{''} (\phi^{(n)}) \ (x-\phi^{(n)})
\ , \label{tangent} \ee
that are tangent to the curve at subsequent points, defined
recursively as
\be \phi^{(n+1)} \ = \ \phi^{(n)} \ - \  \frac{V'
(\phi^{(n)})}{V^{''} (\phi^{(n)})} \ , \ee
where $\phi^{(n)}$ denotes the $n$-th iteration of the wrong
vacuum $\phi^{(0)}= \phi_0$.

When applied to our case, restricting our attention to the first
terms the method gives
\ba \phi^{(1)} &=& \phi_0 - \frac{V'}{V^{''}} \ ,
\nonumber \\
\phi^{(2)} &=& \phi^{(1)} -
\frac{V'(\phi^{(1)})}{V^{''}(\phi^{(1)})}=
 \phi_0 + \frac{V'}{V^{''}} \left[ \, - 1 \ - \ \frac{ \frac{V'\; V^{'''}}{2 (V^{''})^2} - \frac{(V')^2 \;
V^{''''}}{6 (V^{''})^3}}{1 - \frac{V'V^{'''}}{(V^{''})^2} +
\frac{(V')^2
V^{''''}}{2 (V^{''})^3}}\, \right] \nonumber \\
&\cong & \phi_0  + \frac{V'}{V^{''}} \biggl[-1
-\frac{1}{2}(y_1-\frac{y_2}{3})- \frac{1}{2}(y_1-\frac{y_2}{3})
(y_1-\frac{y_2}{2}) \biggr] \ , \label{ft25} \ea
where $y_1$ and $y_2$ are defined in (\ref{ft23}) and, for
brevity, the arguments are omitted whenever they are equal to
$\phi_0$. Notice the precise agreement with the first four terms
in (\ref{ft24}), that imply that our tadpole resummations have a
simple interpretation in terms of successive iterations of the
solutions of the vacuum equations $V'=0$ by Newton's method.
Notice the emergence of the combination $y_1-y_2/3$ after the
first iteration: as a result, the pattern of eqs. (\ref{ft24})
continues indeed to all orders.

In view of this interpretation, the non-renormalization points
$\phi_0 = \pm a/2$ acquire a clear geometrical interpretation: in
these cases the iteration stops after the first term $\phi^{(1)}$,
since the tangent drawn at the original ``wrong'' vacuum, say at
$a/2$, happens to cross the real axis precisely at the extremum on
the other side of the barrier, at $\langle \phi \rangle = -a$.
Newton's method can also shed some light on the behavior of the
iterations, that stay on one side of the extremum or pass to the
other side according to the concavity of the potential, and on the
convergence radius of our tadpole resummations, that the second
iteration already restricts $y_1$ and $y_2$ to the region
\be \left|y_1\ - \ \frac{y_2}{2}\right| \ < \ 1 \ . \label{ft27}
\ee

However, the tangent method behaves as a sort of Dyson resummation
of the naive diagrammatic expansion, and has therefore better
convergence properties. For instance, starting near the
non-renormalization point $\phi_0 \simeq -a/2$, the first
iteration lands far away, but close to the minimum $\phi^{(1)}
\simeq a$. The second correction, that when regarded as a
resummation in (\ref{ft24}) is large, is actually small in the
tangent method, since it is proportional to $(1/16) (y_1-y_2/3)$.
It should be also clear by now that not only the points
$\phi_0=\pm a/2$, but finite intervals around them, move across
the barrier as a result of the iteration. These steps, however, do
not have a direct interpretation in terms of Feynman diagram
tadpole resummations, since (\ref{ft27}) is violated, so that the
corresponding diagrammatic expansion actually diverges. The reason
behind the relative simplicity of the cubic potential (\ref{ft33})
is easily recognized from the point of view of the tangent method:
the corresponding $V'(\phi)$ is a parabola, for which Newton's
method never leads to tangents crossing the real axis past an
inflection point.

\begin{figure}[h]
\begin{center}
\resizebox{5cm}{!}{\psfig{figure=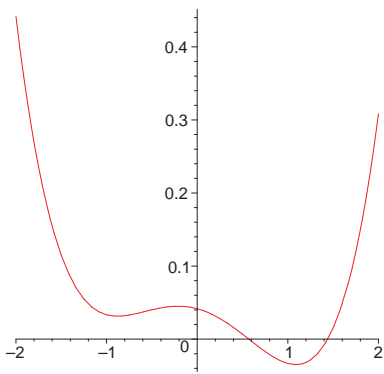,width=5cm}}
\caption{A quartic potential with a ``magnetic'' deformation}
\label{magnetic}
\end{center}
\end{figure}

There is a slight technical advantage in returning to the example
of eq. (\ref{modelaction}), since for a small magnetic field
(tadpole) $c$ (fig. \ref{magnetic}) one can expand the complete
expressions for the vacuum energy and the scalar {\it v.e.v.}'s in
powers of the tadpole. The expansions (\ref{ft24}) still apply,
with an obvious change in the one-point function $V'$, and their
sums should coincide, term by term, with the tadpole resummations
obtained starting from the undeformed ``wrong'' vacua $\phi_0 =
\pm a,0$.

In this case the vacuum energy is given by
\be \Lambda_0 \ = \ \frac{\lambda}{4!}\, (v^2-a^2)^2 \ - \ c \; v
\ , \label{condens} \ee
where the correct vacuum value $\langle\phi\rangle = v $ is
determined by the cubic equation
\be \lambda \frac{v}{6} \; (v^2 - a^2) \ - \ c \ = \ 0 \ ,
\label{cubiceq} \ee
that can be easily solved perturbatively in the tadpole $c$, so
that if one starts around $\phi_0=a$,
\be \Lambda_0 \ = \ - \,  c \; a \ - \ \frac{3 c^2}{2 \lambda a^2}
\ + \ {\cal O}(c^3) \ . \ee
This result can be also recovered rather simply starting from the
wrong vacuum $\phi_0=a$ and making use of eq.~(\ref{lambda1}),
since in this case $\Delta=-c$ and ${\cal D}(p^2=0)=- 3i /\lambda
a^2$. However, the cubic equation (\ref{cubiceq}) can be also
solved exactly in terms of radicals, and in the small tadpole
limit its three solutions are real and can be written in the form
\be v \ = \ \frac{2a}{\sqrt{3}} \ \cos\left[ \frac{\alpha}{3} +
\frac{2 \pi k}{3} \right] \qquad (k=0,1,2) \ , \label{cubicsol}
\ee where \be \cos(\alpha) = \xi \ , \qquad \sin(\alpha) = \sqrt{1
- \xi^2} \ , \quad {\rm with} \quad \xi = \frac{9 \sqrt{3}
c}{\lambda a^3} \ . \ee
For definiteness, let us consider a tadpole $c$ that is small and
positive, so that the absolute minimum of the deformed potential
lies in the vicinity of the original minimum of the Mexican-hat
section at $v=a$ and corresponds to $k=0$. We can now describe the
fate of the resummations that start from two different wrong
vacua:

i) $\phi_0 = a$. In this case and in the small tadpole limit $\xi
<<1$ resummations in the diagrammatic language produce the first
corrections
\be
 \langle\phi\rangle = a + \frac{3 c}{\lambda a^2} -\frac{27 c^2}{2
 \lambda^2 a^5} + \cdots \ . \label{ft7}
\ee
Alternatively, this result could be obtained solving
eq.~(\ref{cubiceq}) in powers of the tadpole $c$ with the initial
value $\phi_0=a$, so that, once more, starting from a wrong vacuum
close to an extremum and resumming one can recover the correct
answer order by order in the expansion parameter. In this case
both a cubic and a quartic vertex are present, and the complete
expression for the vacuum energy, obtained substituting
(\ref{cubicsol}) in (\ref{condens}), is
\be \Lambda_0 \ = \ \frac{2\; \lambda a^4 \; \xi}{27} \ \left[
\frac{\xi}{16} \, \frac{1}{\cos^2\left( \frac{\alpha}{3} \right)}
\ - \ \cos\left( \frac{\alpha}{3} \right)  \right] \ .
\label{clvaen} \ee
This can be readily expanded in a power series in $\xi$, whose
first few terms,
\be \Lambda_0 \ = \ - a \; c \ - \ \frac{\sqrt{3}}{18} \; a \; c
\; \xi \ + \ \frac{1}{54} \; a \; c \; \xi^2 \ - \
\frac{\sqrt{3}}{243} \; a \; c \; \xi^3 \ + \ {\cal O}(\xi^4) \ .
\label{powerser} \ee match precisely the tadpole expansion
(\ref{ft24}).

ii)  $\phi_0 = 0$. In this case the first corrections obtained
resumming tadpole diagrams are
\be
 \langle\phi\rangle = - \frac{6 c}{\lambda a^2} -\frac{216 c^3}{
 \lambda^3 a^8} + \cdots \ . \label{ft8}
\ee
The same result can be obtained expanding the solution of
(\ref{cubiceq}) in powers of the tadpole $c$, starting from the
initial value $\phi_0=0$. We can now compare (\ref{cubicsol}) with
(\ref{ft24}), noting that in this case only the quartic vertex is
present, so that $y_1=0$. Since $y_2 = - 8 \xi^2/9$ for
$\phi_0=0$, the {\it v.e.v.} $\langle \phi \rangle$ contains only
odd powers of the tadpole $c$, a property that clearly holds in
(\ref{cubicsol}) as well, since $\xi \rightarrow - \xi$
corresponds to $\alpha \rightarrow \alpha + (2l+1) 3 \pi$, with
$l$ integer and $v \rightarrow -v$ in (\ref{cubicsol}). The
contributions to $\langle \phi \rangle$ are small and negative,
and therefore, starting from a wrong vacuum close to a maximum of
the theory, the resummation flow leads once more to a nearby
extremum (the local maximum slightly to the left of the origin, in
this case), rather than rolling down to the minimum corresponding
to the $k=0$ solution of (\ref{cubicsol}). The important points in
the scalar potential are again the extrema and the inflections,
precisely as we had seen in the example with a cubic potential.
Barring the peculiar behavior near the points identified by the
condition $y_1=y_2/3$, the scalar field flows in general to the
nearest extremum (minimum or maximum) of this potential, without
passing through any inflection point along the whole resummation
flow. It should be appreciated how the link with Newton's method
associates a neat geometrical interpretation to this behavior.

For $\xi = 1$ the two extrema of the unperturbed potential located
at $ \ v=-a \ $ and at $ \ v=0 \ $ coalesce with the inflection at
$-a/\sqrt{3}$. If the potential is deformed further, increasing
the value of the tadpole, the left minimum disappears and one is
left with only one real solution, corresponding to $k = 0$. The
correct parameterization for $ \ \xi>1 \ $ is $ \
\cosh(\alpha)=\xi \ $ and $ \ \sinh(\alpha) = \sqrt{\xi^2-1} \ $,
and the classical vacuum energy is like in (\ref{clvaen}), but
with $ \ \cos(\alpha/3)$ replaced by $\cosh(\alpha/3)$. In order
to recover the result (\ref{powerser}) working perturbatively in
the ``wrong'' vacuum $v = a$, one should add the contributions of
an infinite series of diagrams that build a power series in
$1/\xi$, but this cannot be regarded as a tadpole resummation
anymore. The meaning of the parameter $\xi$ should by now have
become apparent: it is proportional to the product of the tadpole
and the propagator in the wrong vacuum, $V'/V''$, a natural
expansion parameter for problems of this type. Notice that the
ratio $V'/V^{''}$ is twice as large (and of opposite sign) at the
origin $\phi_0=0$ than at $\phi_0=a$. Therefore, the tadpole
expansion first breaks down around $\phi_0=0$. As we have seen,
the endpoint of the resummation flow for $\phi_0=0$ is the local
maximum corresponding to $k=2$ in (\ref{cubicsol}), that is
reached for $\xi < 1$. At $\xi=1$, however, the two extrema
corresponding to $k=1$ and $k=2$ coalesce with the inflection at
$-a/\sqrt{3}$, and hence there is no possible endpoint for the
resummation flow. This is transparent in (\ref{cubicsol}), since
for $\xi \ge 1$ the $\xi$ expansion clearly breaks down. It should
therefore be clear why, if the potential is deformed too
extensively, corresponding to $ \ \xi >1$, a perturbative
expansion around the extrema $ \ v= \pm a ,0\ $ in powers of $\xi$
is no longer possible. Another key issue that should have emerged
from this discussion is the need for an independent, small
expansion parameter when tadpoles are to be treated
perturbatively. In this example, as anticipated, the expansion
parameter $\xi$ can be simply related to the potential according
to
\be \xi \ = \ \frac{3 \sqrt{3}}{a} \ \left| \frac{V'}{V''}
\right| \ , \ee
that is indeed small if $c$ is small.

It is natural to ask about the fate of the points $\phi_0 = \pm
a/2$ of the previous example (\ref{phitothefour}) when the
magnetic field $c$ is turned on. Using the parameters $y_i$ in
(\ref{ft23}), the condition determining these special points,
$y_1=y_2/3$, is equivalent to
\be V' \ = \ 3 \ \phi_0 \ V^{''} \ . \label{ft99} \ee
It is readily seen that the solutions of this cubic equation are
precisely $\phi_0 = -v /2$, if $v$ denotes, collectively, the
three extrema solving (\ref{cubiceq}). Hence, in this case
$\langle \phi \rangle = - 2 \phi_0 = v$, confirming the
persistence of these non-renormalization points. In the present
case, however, there is a third non-renormalization point, $\phi_0
= - v_{k=2} /2$, that for small values of $\xi$ is well inside the
convergence region. This last point clearly admits an
interpretation in terms of tadpole resummations, and the possible
existence of effects of this type in String Theory raises the hope
that explicit vacuum redefinitions could be constructed in a few
steps for special values of the string coupling and of other
moduli.

\subsection{Branes and tadpoles of codimension one}

Models whose tadpoles are confined to lower-dimensional surfaces
are of particular interest. In String Theory there are large
classes of examples of this type, including brane supersymmetry
breaking models \cite{sugimoto,bsb}, intersecting brane models
\cite{intersecting} and models with internal fluxes \cite{fluxes}.
If the space transverse to the branes is large, the tadpoles are
"diluted" and there is a concrete hope that their corrections to
brane observables be small, as anticipated in \cite{bsb}. In the
codimension one case, tadpoles reflect themselves in boundary
conditions on the scalar (dilaton) field and hence on its
propagator, and as a result their effects on the Kaluza-Klein
spectrum and on brane-bulk couplings are nicely tractable.

Let us proceed by considering again simple toy models that display
the basic features of lower-dimensional tadpoles. The internal
spacetime is taken to be $S^1/Z_2$, with $S^1$ a circle, and the
coordinate of the circle is denoted by $y$: in a string
realization its two endpoints $y = 0$ and $y = \pi R $ would be
the two fixed points of the orientifold operation $\Omega' =
\Omega \ \Pi_y $, with $\Pi_y$ the parity in $y$. We also let the
scalar field interact with a boundary gauge field, so that
\be S = \int d^4 {\bf x} \int_0^{\pi R} d y \biggl\{ \
-\frac{1}{2} ({\partial \phi})^2 \ - \ \left(T \ \phi \ + \
\frac{m}{2} \ \phi^2 \ - \ \phi \ \tr(F^2)\right) \ \delta (y -
\pi R) \ \biggr\} \ . \label{l9} \ee
The Lagrangian of this toy model describes a free massless scalar
field living in the bulk, but with a tadpole and a mass-like term
localized at one end of the interval $[0,\pi R]$. In String
Theory, both the mass-like parameter $m$ and the tadpole $T$ in
the examples we shall discuss would be perturbative in the string
coupling constant $g_s$. Any non-analytic IR behavior associated
with the possible emergence of $1/m$ terms would thus signal a
breakdown of perturbation theory, according to the discussion
presented in the Introduction. Notice that, for dimensional
reasons, the mass term is proportional to $m$, rather than to
$m^2$ as is usually the case for bulk masses.

The starting point is the Kaluza-Klein expansion
\be
 \phi ({\bf x},y) \ = \ \phi_c (y) + \sum_k \chi_k (y) \ \phi_k ({\bf x}) \
 , \label{l3}
\ee
where $\phi_c (y)$ is the classical field and the $\phi_k ({\bf
x})$ are higher Kaluza-Klein modes. The classical field $ \ \phi_c
\ = \ - \ T/m \ $ solves the simple differential equation
\be
\phi_c^{''} \ = \ 0 \  \ee
in the internal space, with the boundary conditions
\ba
&& \phi_c^{'} \ = \ 0 \  \qquad {\rm at} \ y = 0 \ , \nonumber\\
&& \phi_c^{'} \ = \ - \ T \ - \ m \ \phi_c  \ \qquad {\rm at} \ y
= \pi R \ , \label{l10} \ea
while the Kaluza-Klein modes satisfy in the internal space the
equations
\be \chi_k^{''} \ + \ M_k^2 \ \chi_k \ = \ 0 \ , \ee
with the boundary conditions
\ba &&  \chi_k^{'} \ = \ 0 \quad{\rm at} \quad y \ = \ 0 \ , \nonumber \\
&&  \chi_k^{'} \ = \ - \ m \ \chi_k  \quad {\rm at} \quad {\rm at}
\ y \ = \ \pi R \ . \label{l11} \ea
The corresponding solutions are then
\be
\chi_k(y) \ = \ A_k \ \cos{(M_k y)} \ , \label{l12}
\ee
where the masses $M_k$ of the Kaluza-Klein modes are determined by the eigenvalue equation
\be
M_k\tan{(M_k\pi R)} \ = \ m \ . \label{l13}
\ee

The classical vacuum energy can be computed directly working in
the ``right'' vacuum. To this end, one ignores the Kaluza-Klein
fluctuations and evaluates the classical action in the correct
vacuum, as determined by the zero mode, with the end result that
\be \Lambda_0 \ = \ - \ \frac{T^2}{2m} \ . \label{l14} \ee
One can similarly compute in the ``right'' vacuum the gauge coupling, obtaining
\be
\frac{1}{g^2} \ = \ - \ \phi_c(y=\pi R) \ = \ \frac{T}{m} \ .
\label{l16}
\ee

It is amusing and instructive to recover these results expanding
$\phi$ around the ``wrong'' vacuum corresponding to vanishing
values for both $T$ and $m$. The Kaluza-Klein expansion is
determined in this case by the Fourier decomposition
\be \phi ({\bf
x}, y) = \frac{1}{\sqrt{\pi R}} \ \sum_{k=0}^{\infty}  b_k \cos
\left(\frac{k y}{R}\right) \ \phi^{(k)} ({\bf x}) \ , \
\label{l17} \ee
where $b_0=1$ and $b_k= \sqrt{2}$ for $k \neq 0$,
that turns the action into
\be S_{KK} \ = \ \int d^4 {\bf x} \left\{\, -\,
\frac{1}{2}\sum_{k,l\geq 0} \phi_k\; \mathcal{M}^2_{k,l}\; \phi_l
\ - \ T\sum_{k\geq 0} \frac{b_k(-)^k}{\sqrt{\pi R}}\,
\phi_k\right\} \ . \label{l32} \ee
Here we are ignoring the kinetic term, since the vacuum energy is
determined by the zero momentum propagator, while the mass matrix
is
\be {\mathcal M}^2_{kl} \ = \
\frac{k^2}{R^2}\ \delta_{k,l} \ + \ \frac{m}{\pi R} \ b_k b_l \
(-)^{k+l} \ . \label{l33} \ee

The eigenvalues of this infinite dimensional matrix can be
computed explicitly using the techniques in \cite{ddg}. It is
actually a nice exercise to show that the characteristic equation
defining the eigenvalues of (\ref{l33}) is precisely (\ref{l13}),
and consequently that the eigenvectors of (\ref{l33}) are the
fields $\chi_k$ defined in (\ref{l12}). In fact, multiplying
(\ref{l33}) by normalized eigenfunctions $\Psi^{\lambda}_k$ gives
\be
\Psi^{\lambda}_k \ = \ (-)^k \ \frac{m b_k}{\pi R} \ \frac{
\sum_{l}b_l(-)^l\Psi^{\lambda}_l}{\lambda^2 - \frac{k^2}{R^2}} \ ,
\label{l35} \ee
so that
\be \langle k|\lambda\rangle \ = \ \Psi_k^{\lambda} \ = \
\mathcal{N}_{\lambda} \frac{b_k(-)^k}{\frac{k^2}{R^2}-\lambda^2} \
. \label{l37} \ee
 and therefore
\be \sum_{k=0}^{\infty} \frac{b_k^2}{\lambda^2 - \frac{k^2}{R^2}}
\ = \ \frac{\pi R}{m} \ . \label{l36} \ee
The sum can be related to a well-known representation of
trigonometric functions \cite{ww},
\be {\rm cotg}(\lambda \pi R) \ = \ \frac{\lambda}{\pi R} \
\sum_{k=0}^\infty \frac{b_k^2}{\lambda^2 - \frac{k^2}{R^2}} \ ,
\ee
and hence the eigenvalues of (\ref{l33}) coincide with those of
(\ref{l13}). In order to compute the vacuum energy, one needs in
addition the k-component of the eigenvector $|\lambda\rangle$,
that can be read from (\ref{l35}). The normalization constant
$\mathcal{N}_{\lambda}$ in (\ref{l37}) is then determined by the
condition
\be 1 \ = \ \langle \lambda|\lambda\rangle \ = \
\mathcal{N}_{\lambda}^2\
\sum_{k=0}\frac{b_k^2}{\left(\frac{k^2}{R^2}-\lambda^2\right)^2} \
= \ \frac{\mathcal{N}_{\lambda}^2}{2\lambda} \ \frac{d}{d \lambda}
\ \sum_{k=0}\frac{b_k^2}{\frac{k^2}{R^2}-\lambda^2} \ ,
\label{l38} \ee
that using again eq. (\ref{l13}) can be put in the form
\be
\mathcal{N}^2_\lambda \ = \ \frac{2m^2\lambda^2}{\pi^2 R^2} \
\frac{1}{\lambda^2+\alpha^2} \ , \label{l39} \ee
with
\be
\alpha^2 = \frac{m}{\pi R} \left( 1 + \pi R \; m \right) \ . \label{alpha2}
\ee

Notice that in the limit $R \, m <<1$, that in a string context,
where $m$ would be proportional to the string coupling, would
correspond to the small coupling limit \cite{pw}, the physical
masses in (\ref{l13}) are approximately determined by the
solutions of the linearized eigenvalue equation, so that
\ba
M_0^2 &\cong& \frac{m}{\pi R} \ , \nonumber \\
M_k^2 &\cong& \frac{k^2}{R^2} + 2 \frac{m}{\pi R} \ . \label{l039}
\ea

One can now recover the classical vacuum energy using eq.
(\ref{lambda1}),
\be \Lambda_0 \ = \ -\frac{i}{2}\, \sum_{k,l}\Delta^{(k)}(y_1 =
\pi R) \, \langle k |{\cal D} ({\bf 0}~; y_1=\pi R,y_2=\pi R) |l
\rangle \, \Delta^{(l)}l(y_2 = \pi R) \ , \label{l45} \ee
that after inserting complete sets of eigenstates becomes
\be \Lambda_0 \ = \ -\, \frac{T^2}{2 \pi R}\sum_{\lambda, k,l} \
b_k b_l(-)^{k+l} \ \langle k|\Psi_\lambda\rangle \
\frac{1}{\lambda^2} \ \langle \Psi_\lambda|l\rangle \ = \
-\frac{T^2}{2 \pi R}\sum_{\lambda} \
\frac{\mathcal{N}_\lambda^2}{\lambda^2}
 \ \left(\sum_{k=0} \frac{b_k^2}{\frac{k^2}{R^2}-\lambda^2}\right)^2 \ ,
\label{l40} \ee
or, equivalently, using eq. (\ref{l36})
\be \Lambda_0 \ = \ -\, \frac{T^2}{\pi R} \
\sum_{\lambda}\frac{1}{\lambda^2+\alpha^2} \ . \label{l41} \ee

The sum over the eigenvalues in (\ref{l41}) can be finally
computed by a Sommerfeld-Watson transformation, turning it into a
Cauchy integral according to
\be
\sum_{\lambda}\frac{1}{\lambda^2+\alpha^2} \ = \ \frac{1}{2} \
\oint\frac{d z}{2\pi i} \ \frac{1}{z^2+\alpha^2} \ \frac{(1+m\pi
R)\sin(\pi R z)+\pi R z\cos(\pi R z)}{z\sin(\pi R z)-m\cos(\pi R
z)} \ . \label{l42} \ee
The path of integration encircles the real axis, but can be
deformed to contain only the two poles at $ z= \pm i\alpha $. The
sum of the corresponding residues reproduces again (\ref{l14}),
since
\be
\frac{1}{2\alpha} \ \frac{(1+m\pi R)\sinh(\pi R \alpha)+\pi R
\alpha\cosh(\pi R \alpha)} {\alpha\sinh(\pi R \alpha)+m\cosh(\pi R
\alpha)} \ = \ \frac{\pi R}{2 m} \ , \label{l43}
\ee
where we used the definition of $\alpha$ in (\ref{alpha2}), even
though the computation was now effected starting from a wrong
vacuum.

It is useful to sort out the contributions to the vacuum energy
coming from the zero mode $\Lambda_0^{(0)}$ and from the massive
modes $\Lambda_0^{(m)}$. In a perturbative expansion using eq.
(\ref{l039}), one finds
\ba \Lambda_0^{(0)} &\cong&  - \ \frac{T^2}{2m} \ + \
\frac{T^2}{4}\ \pi R \ ,
\nonumber \\
\Lambda_0^{(m)} &\cong& - \ \frac{T^2}{4} \ \pi R \ . \label{l47}
\ea

Notice that the correct result (\ref{l14}) for the classical
vacuum energy is completely determined by the zero mode
contribution, while to leading order the massive modes simply
compensate the perturbation introduced by the tadpole, that in
String Theory would be interpreted, by open-closed duality, as the
one-loop gauge contribution to the vacuum energy. In a similar
fashion, in the wrong vacuum the gauge coupling can be read simply
from the amplitude with two external background gauge fields going
into a dilaton tadpole. In this case there are no other
corrections with internal gauge lines, since we are only
considering a background gauge field. The result for the gauge
couplings is then
\be
\frac{1}{g^2} \ = \ \frac{T}{\pi R}\sum_{\lambda, k,l} \ b_k b_l(-)^{k+l}
\ \langle k|\Psi_\lambda\rangle \ \frac{1}{\lambda^2} \ \langle \Psi_\lambda|l\rangle
\ = \ \frac{T}{\pi R}\sum_{\lambda} \ \frac{\mathcal{N}_\lambda^2}{\lambda^2}
 \ \left(\sum_{k=0} \frac{b_k^2}{\frac{k^2}{R^2}-\lambda^2}\right)^2 \ ,
\ee
so that using eq.~(\ref{l39}) for $ \ \mathcal{N}^2_\lambda \ $
and performing the sum as above one can again recover the correct
answer, displayed in (\ref{l16}).

\subsection{Branes and tadpoles of higher codimension}

Antoniadis and Bachas argued that in orientifold models the
quantum corrections to brane observables \cite{ab} have a
negligible dependence on the moduli of the transverse space for
codimension larger than two. This result is due to the rapid
falloff of the Green function in the transverse space, but rests
crucially on the condition that the global $NS$-$NS$ tadpole
conditions be fulfilled. In this Section we would like to
generalize the analysis to models with $NS$-$NS$ tadpoles,
investigating in particular the sensitivity to scalar tadpoles of
the quantum corrections to brane observables. To this end, let us
begin by generalizing to higher codimension the example of the
previous Subsection, with
\be S = \int d^4 {\bf x} \int_0^{\pi R} d^n y \left\{ \
-\frac{1}{2} ({\partial \phi})^2 \ - \ \left(T \ \phi \ + \
\frac{m^2}{2} \ \phi^2\right) \ \delta^{(n)} (y)  \right\} \ .
\label{h1} \ee
The correct vacuum and the correct classical vacuum energy in this
example are clearly
\be
 \phi_c \ = \ - \, \frac{T}{m^2} \ , \qquad
\Lambda_0 \ = \ - \ \frac{T^2}{2m^2} \ . \label{h2} \ee
For simplicity, we are considering a symmetric compact space of
volume $V_n \equiv (\pi R)^n$, so that the Kaluza-Klein expansion
in the wrong vacuum is
\be \phi ({\bf x}, y) = \frac{1}{\sqrt{V_n}} \ \sum_{{\bf k}}
\prod_{i=1}^n \left[ b_{ k_i} \cos \left(\frac{k_i y_i}{R}\right)
\right] \ \phi^{({\bf k})} ({\bf x}) \ . \ \label{h3} \ee
After the expansion, the action reads
\be S_{KK} \ = \ \int d^4 {\bf x} \left( -\frac{1}{2}\sum_{{\bf
k},{\bf l}} \phi_{\bf k} \, \mathcal{M}^2_{{\bf k},{\bf l}}\,
\phi_{\bf l} \ - \ T\sum_{{\bf k}} \frac{b_{\bf k}}{\sqrt{V_n}}\
\phi_{\bf k} \right) \ , \label{h4} \ee
where, as in the previous Section, we neglected the space-time
kinetic term, that does not contribute, and where the mass matrix
is
\be {\mathcal M}^2_{{\bf k} {\bf
l}} \ = \ \frac{{\bf
    k}^2}{R^2}\ \delta_{{\bf k},{\bf l}} \ + \ \frac{m^2}{V_n} \ b_{\bf
  k} b_{\bf l} \ . \label{h5}
\ee
In this case the physical Kaluza-Klein spectrum is determined by
the eigenvalues $\lambda$ of the mass matrix (\ref{h5}), and hence
is governed by the solutions of the ``gap equation''
\be 1 \ = \ \frac{m^2}{V_n} \sum_{\bf k} \frac{b_{\bf
k}^2}{\lambda^2 - {{\bf k}^2 \over
    R^2}} \ . \label{h6}
\ee

We thus face a typical problem for Field Theory in all cases of
higher codimension, the emergence of ultraviolet divergences in
sums over bulk Kaluza-Klein states. In String Theory these
divergences are generically cut off\footnote{The real situation is
actually more subtle. These divergences are infrared divergences
from the dual, gauge theory point of view, and are not regulated
by String Theory \cite{dm1}. However, this subtlety does not
affect the basic results of this Section.} at the string scale
$|{\bf k}| < R M_s$, and in the following we shall adopt this
cutoff procedure in all UV dominated sums. In the small tadpole
limit $R m << 1$, approximate solutions to the eigenvalue equation
can be obtained, to lowest order, inserting the Kaluza-Klein
expansion (\ref{h3}) in the action, while the first correction to
the masses of the lightest modes can be obtained integrating out,
via tree-level diagrams, the heavy Kaluza-Klein states. In doing
this, one finds that to first order the physical masses are given
by
\ba M_0^2 &=& \frac{m^2}{V_n} \ \left(1 + c \, \frac{m^2}{M_s^2} +
\cdots \right) \ ,
\nonumber \\
M_k^2 &=& \frac{k^2}{R^2} + 2 \, \frac{m^2}{V_n} + \cdots \ .
\label{h7} \ea
The would-be zero mode thus acquires a small mass that, as in the
codimension-one example, signals a breakdown of perturbation
theory, whereas the corrections to the higher Kaluza-Klein masses
are very small and irrelevant for any practical purposes. We would
like to stress that the correct classical vacuum energy (\ref{h2})
is precisely reproduced, in the wrong vacuum, by the
boundary-to-boundary propagation of the single lightest mode,
since
\be \Lambda_0 \ = \ - \, \frac{1}{2} \ \frac{T}{\sqrt{V_n}} \,
\left(\frac{m^2}{V_n}\right)^{-1} \frac{T}{\sqrt{V_n}} \ ,
\label{h8} \ee
while the breaking of string perturbation theory is again manifest
in the nonanalytic behavior as $m \to 0$, so that the contribution
(\ref{h8}) is actually classical. On the other hand, as expected,
the massive modes give contributions that would not spoil
perturbation theory and that, by open-closed duality, in String
Theory could be interpreted as brane quantum corrections to the
vacuum energy. This conclusion is valid for any brane observable,
and for instance can be explicitly checked in this example for the
gauge couplings. This strongly suggests that for quantities like
differences of gauge couplings for different gauge group factors
that, to lowest order, do not directly feel the dilaton zero mode,
quantum corrections should decouple from the moduli of the
transverse space, as advocated in \cite{ab}. The main effect of
the tadpoles is then to renormalize the tree-level (disk) value,
while the resulting quantum corrections decouple as in their
absence.

\subsection{On the inclusion of gravity}

The inclusion of gravity, that in the Einstein frame enters the
low-energy effective field theory of strings via
\be S \ = \ \frac{1}{2 k^2} \, \int d^{\cal D} x \; \sqrt{-g}
\left( R - \frac{1}{2} (\partial \varphi)^2 \right) \ ,
\label{gravity} \ee
presents further subtleties.
First, one is dealing with a gauge
theory, and the dilaton tadpole
\be \delta S \ = \ - T  \int d^{\cal D} x\ \sqrt{-g} \ e^{b
\varphi} \ , \label{gravity2} \ee
when developed in a power series around
the wrong Minkowski vacuum according to
\be g_{\mu\nu}\ = \ \eta_{\mu\nu} \ + \ 2 \;k\; h_{\mu\nu} \ ,
\qquad \varphi = \varphi_0 \ + \ \sqrt{2} \; k \; \phi \ , \ee
appears to destroy the gauge symmetry. For instance,
up to quadratic order it results in tadpoles, masses and mixings
between dilaton and graviton, since
\be T \sqrt{-g} \ e^{b \varphi} \ = \ T e^{b \varphi_0} \left[ 1 +
k h + b \phi - k^2 \left( h_{\mu\nu} h^{\mu\nu} - \frac{1}{2} h^2
\right) + k b \phi h + \frac{b^2}{2} \phi^2 + \ldots \right] \ ,
\ee
where $h$ denotes the trace of $h_{\mu\nu}$. If these terms were
treated directly to define the graviton propagator, no gauge
fixing would seem to be needed. On the other hand, since the fully
non-linear theory does possess the gauge symmetry, one should
rather insist and gauge fix the Lagrangian as in the absence of
the tadpole. Even when this is done, however, the resulting
propagators present a further peculiarity, that is already seen
ignoring the dilaton: the mass-like term for the graviton is not
of Fierz-Pauli type, so that no van Dam-Veltman-Zakharov
discontinuity \cite{vdvz} is present and a ghost propagates.
Finally, the mass term is in fact tachyonic for positive tension,
the case of direct relevance for brane supersymmetry breaking, a
feature that can be regarded as a further indication of the
instability of the Minkowski vacuum.

All these problems notwithstanding, in the spirit of this work it
is reasonable to explore some of these features referring to a toy
model, that allows to cast the problem in a perturbative setting.
This is obtained coupling the linearized Einstein theory with a
scalar field, adding to the Lagrangian (\ref{gravity})
\be \delta {\cal L} \ = \  - \frac{\lambda}{4!} (\phi^2 - a^2)^2 +
\frac{m^2}{2} (h_{\mu\nu} h^{\mu\nu} - h^2) + (\phi^2 - a^2) (h^2
+ b h) \ . \ee

This model embodies a couple of amusing features: in the correct
vacuum $\langle \phi \rangle = a$, the graviton mass is of
Fierz-Pauli type and describes five degrees of freedom in ${\cal
D}=4$, the vacuum energy vanishes, and no mixing is present
between graviton and dilaton. On the other hand, in the wrong
vacuum $\langle \phi \rangle = a(1+ \epsilon)$, the expected
${\cal O}(\epsilon)$ tadpoles are accompanied not only by a vacuum
energy
\be \Lambda_0 \ = \  \frac{\lambda a^4}{6} \, \epsilon^2   \ ,
\label{wrongvacen} \ee
but also by a mixing between $h$ and $\phi$ and by an ${\cal O}(\epsilon)$
modification of the graviton mass, so that to quadratic order
\be \delta {\cal L} \to   - \frac{\lambda}{4!} (4 \epsilon^2 a^4 +
8 a^3 \epsilon \phi + 4 a^2 \phi^2 ) + \frac{m^2}{2} (h_{\mu\nu}
h^{\mu\nu} - h^2) + 2 \epsilon a^2 (b h + h^2) + 2 a b \phi h \ .
\ee
Hence, in this model the innocent-looking displacement to the
wrong vacuum actually affects the degrees of freedom described by
the gravity field, since the perturbed mass term is no more of
Fierz-Pauli type. It is instructive to compute the first
contributions to the vacuum energy starting from the wrong vacuum.
To this end, one only needs the propagators for the tensor and
scalar modes at zero momentum to lowest order in $\epsilon$,
\ba && \langle h_{\mu\nu} h_{\rho\sigma} \rangle_{k=0}\ = \
\frac{i}{2 m^2} \left[ \eta_{\mu\rho} \eta_{\nu\sigma} +
\eta_{\mu\sigma} \eta_{\nu\rho} - \frac{4 a b - \frac{\lambda a
m^2}{3b}}{2 a b {\cal D} + \frac{\lambda a m^2(1-{\cal D})}{6b}}
\eta_{\mu\nu}\eta_{\rho\sigma} \right]
\ , \nonumber \\
&& \langle h_{\mu\nu} \varphi \rangle_{k=0} \ =\ \frac{i\,
\eta_{\mu\nu}}{2 a b {\cal D} +
\frac{\lambda a m^2(1-{\cal D})}{6b}} \ , \nonumber \\
&& \langle \varphi \varphi \rangle_{k=0} = - \frac{(1-{\cal
D})m^2}{2 a b} \, \frac{i}{2 a b {\cal D} + \frac{\lambda a
m^2(1-{\cal D})}{6b}} \ . \ea

There are three ${\cal O}(\epsilon^2)$ corrections to the vacuum energy,
\ba && {\rm a: \quad} \frac{\lambda b \epsilon^2 a^4 {\cal D}}{3}
\ \frac{1}{2 b {\cal D} +
\frac{\lambda m^2}{6b}(1-{\cal D})} \ , \\
&& {\rm b: \quad} - \ \frac{2}{3} \lambda b \epsilon^2 a^4 {\cal
D} \
\frac{1}{2 b {\cal D} + \frac{\lambda m^2}{6b}(1-{\cal D})} \ , \\
&& {\rm c: \quad} - \ \frac{1}{36 b} \lambda^2 \epsilon^2 a^4
(1-{\cal D}) m^2 \ \frac{1}{2 b {\cal D} + \frac{\lambda
m^2}{6b}(1-{\cal D})} \ , \ea
coming from tensor-tensor, tensor-scalar and scalar-scalar
exchanges, according to eq. (\ref{lambda1}), and their sum is seen
by inspection to cancel the contribution from the initial wrong
choice of vacuum. Of course, there are also infinitely many
contributions that must cancel, order by order in $\epsilon$, and
we have verified explicitly that this is indeed the case to ${\cal
O}(\epsilon^3)$. The lesson, once more, is that starting from a
wrong vacuum for which the natural expansion parameter $\left|
V'/V'' \right|$ is small, one can recover nicely the correct
vacuum energy, even if there is a ghost field in the gravity
sector, as is the case in String Theory after the emergence of a
dilaton tadpole if one insists on quantizing the theory in the
wrong Minkowski vacuum.
\section{On String Theory with tadpoles}
\subsection{Evidence for a new link between string vacua}

We have already stressed that supersymmetry breaking in String
Theory is generally expected to destabilize the Minkowski vacuum
\cite{fs}, curving the background space time. Since the
quantization of strings in curved backgrounds is a notoriously
difficult problem, it should not come as a surprise that little
progress has been made on the issue over the years. There are some
selected instances, however, where something can be said, and we
would like to begin this Section by discussing a notable example
to this effect.

Classical solutions of the low-energy effective action are a
natural starting point in the search for vacuum redefinitions, and
their indications can be even of quantitative value whenever the
typical curvature scales of the problem are well larger than the
string scale and the string coupling is small throughout the
resulting space time. If the configurations thus identified have
an explicit string realization, one can do even better, since the
key problem of vacuum redefinitions can then be explored at the
full string level. Our starting point are some intriguingly simple
classical configurations found in \cite{dmt}. As we shall see,
these solutions allow one to control to some extent vacuum
redefinitions at the string level in an interesting case, a
circumstance of clear interest to gain new insights into String
Theory.

Let us therefore consider the type-I' string theory, the T-dual
version of the type-I theory, with $D8/O8$ brane/orientifold
systems that we shall describe shortly, where for simplicity all
branes are placed at the end points $y=0$ and $y = \pi R$ of the
interval $S^1/Z_2$, the fixed points of the orientifold operation.
Let us also denote by $T_0$ ($q_0$) the tension ($R-R$ charge) of
the $D8/O8$ collection at the origin, and by $T_1$ ($q_1$) the
tension ($R-R$ charge) of the $D8/O8$ collection at the other
endpoint $y = \pi R$. The low-energy effective action for this
system then reads
\ba S &=& {1 \over 2{\kappa}^2} \int d^{10} x \, \sqrt {-G} \left[
e^{-2 \varphi} ( R + 4 ({\partial \varphi})^2)- {1 \over 2 \times
10 !} F_{10}^2
\right] \nonumber \\
&-& \int_{y=0} d^9 x \left(T_0 \ \sqrt{-\gamma} \ e^{-\varphi} \ +
q_0 \ A_9\right) - \int_{y= \pi R} d^9 x  \left(T_1 \sqrt{-\gamma}
\ e^{-\varphi} \ +q_1 A_9 \right) \ , \label{t1b} \ea
where we have included all lowest-order contributions. If
supersymmetry is broken, it was shown in \cite{dmt} that no
classical solutions exist that depend only on the transverse $y$
coordinate, a result to be contrasted with the well-known
supersymmetric case discussed by Polchinski and Witten in
\cite{pw}, where such solutions played an important role in
identifying the meaning of local tadpole cancellation. It is
therefore natural to inquire under what conditions warped
solutions can be found that depend on $y$ and on a single
additional spatial coordinate $z$, and to this end in the Einstein
frame one can start from the ansatz
\be ds^2=e^{2A(y,z)} \eta_{\mu\nu}dx^\mu dx^\nu \ + \ e^{2B(y,z)}
(dz^2+dy^2) \ . \ee

If the functions $A$ and $B$ and the dilaton $\varphi$ are allowed
to depend on $y$ and on $z$, the boundary conditions at the two
endpoints $0$ and $\pi R$ of the interval imply the two
inequalities \cite{dmt}
\be T_0^2 \leq q_0^2  \ , \quad  T_1^2 \leq q_1^2  \ ,
\label{condi2} \ee
necessary but not sufficient in general to guarantee that a
solution exist. As shown in \cite{dmt}, the actual solution
depends on two parameters, $\lambda$ and $\omega$, that can be
related to the boundary conditions at $y=0$ and $y = \pi R$
according to
\be \cos (\omega)= - T_0  / |q_0| \ , \quad \cos (\pi \lambda R +
\omega)=  T_1  / |q_1| \ , \label{t23} \ee
and reads
\ba e^{24 A} &=& e^{b_0+5 \varphi_0/4} \biggl[G_0 + {{3 {\kappa}^2
|q_0|} \over {2 \lambda}} \, e^{
\lambda z} \sin (\lambda |y|+\omega) \biggr] \ , \nonumber \\
e^{24 B} &=&   e^{24 \lambda z +25 b_0+5 \varphi_0/4} \biggl[G_0+
{{3 {\kappa}^2 |q_0|} \over {2 \lambda}} \, e^{
\lambda z } \sin (\lambda |y|+\omega) \biggr] \ , \nonumber \\
e^{\Phi} &=& e^{-5 b_0/6 - \varphi_0/24} \biggl[G_0 + {{3
{\kappa}^2 |q_0|} \over {2 \lambda}} \, e^{ \lambda z } \sin
(\lambda |y|+\omega) \biggr]^{-{5 \over 6}}
 \ , \label{t24}
\ea
where $b_0$, $\varphi_0$ and $G_0$ are integration constants. The
$z$ coordinate is noncompact, and as a result the effective Planck
mass is infinite in this background. There are singularities for
$z \to \pm \infty$ and, depending on the sign of $G_0$ and on the
numerical values of $\lambda$ and $\omega$, the solution may
develop additional singularities at a finite distance from the
origin in the $(y,z)$ plane.

This solution can be actually related to the supersymmetric
solution of \cite{pw}. Indeed, the conformal change of coordinates
\be Y \ = \ {1 \over \lambda} \ e^{\lambda z} \, \sin (\lambda y +
\omega) \ , \quad Z \ = \ {1 \over \lambda} \ e^{\lambda z} \,
\cos (\lambda y + \omega)  \ , \label{t025} \ee
or, more concisely
\be Z+ i~Y \ = \ {e^{i \omega} \over \lambda} \ e^{\lambda (z + i~
y)} \ , \label{map} \ee
maps the strip in the $(y,z)$ plane between the two $O$ planes
into a wedge in the $(Y,Z)$ plane and yields for $y >0$ the
spacetime metric
\be ds^2 = \left[ G_0+ \frac{3{\kappa}^2 |q_0|}{2}\, Y \right]^{1
\over 12} \, \biggl( \eta_{\mu \nu} \, dx^{\mu} \, dx^{\nu} +dY^2+
dZ^2 \biggr) \ . \label{t026} \ee
\begin{figure}[h]
\begin{center}
\resizebox{6cm}{!}{\psfig{figure=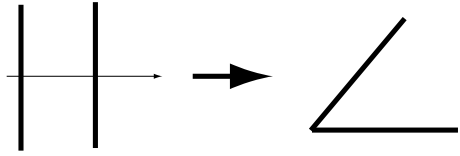,width=6cm}}
\caption{Eq.~(\ref{map}) maps a strip in the $(y,z)$ plane to a
wedge in the $(Y,Z)$ plane} \label{conformal}
\end{center}
\end{figure}

Notice that (\ref{t026}) is the metric derived by Polchinski and
Witten \cite{pw} in the supersymmetric case, but for one notable
difference: here the $Y$ direction is not compact. On the other
hand, in the new coordinate system $(Y,Z)$ the periodicity under
$y \to y+2 \pi R$ reflects itself in the orbifold identification
\be  Z+ i\, Y \ \to \ e^{2\pi i \lambda R} \ (Z+ i\, Y) \ ,
\label{t027} \ee
a two-dimensional rotation ${\cal R}_{\theta}$ in the $(Y,Z)$
plane by an angle $\theta = 2 \pi \lambda R$. In addition, the
orientifold identification $y\rightarrow -y$ maps into a parity
$\Pi_Y$ times a rotation ${\cal R}_{2 \omega}$ by an angle $2
\omega$, so that the new $\Omega$ projection is
\be \Omega' \ = \ \Omega \ \Pi_Y \ {\cal R}_{2 \omega} \ ,
\label{t028} \ee
where $\Omega$ denotes the conventional world-sheet parity. Notice
that both the metric and the dilaton in (\ref{t24}) depend
effectively on the real part of an analytic function, and thus
generally on a pair of real variables, aside from the case of
\cite{pw}, where the function is a linear one, so that one of the
real variables actually disappears. This simple observation
explains the special role of the single-variable solution of
\cite{pw} in this context.

As sketched in fig.~\ref{conformal}, the exponential mapping turns
the region delimited by the two parallel fixed lines of the
orientifold operations in the $(y,z)$ plane into a wedge in the
$(Y,Z)$ plane, delimited by the two lines
\ba
\Omega'~: \  Y &=& \tan \omega \ Z \ , \nonumber \\
\Omega'~ {\cal R}_{2 \pi \lambda R} : \  Y &=& \tan (\pi \lambda R
+ \omega ) \ Z \ , \label{t029} \ea
so that the orientifolds and the branes at $y=0$ form an angle
$\theta_0 = \ \omega $ with the $Z$ axis, while those at $y=\pi R$
form an angle $ \theta_1 = \ \pi\lambda R + \omega $. Notice that
the orbifold identification (\ref{t027}) implies that in general
the two-dimensional $(Y,Z)$ plane contains singularities. In order
to avoid subtleties of this type, in what follows we restrict our
attention to a case where this complication is absent.

The example we have in mind is a variant of the M-theory breaking
model of \cite{ads1}. Its oriented closed part is related by a
T-duality to a Scherk-Schwarz deformation of the toroidally
compactified IIB spectrum of \cite{ads1} (our conventions are
spelled out in the reviews in \cite{orientifolds}), described by
\ba \mathcal{T} &=& ( |V_8|^2+|S_8|^2) \Lambda_{m,2n}  + (
|O_8|^2+|C_8|^2) \Lambda_{m,2n+1}
\nonumber\\
&-& ( V_8\bar{S}_8+S_8\bar{V}_8 ) \Lambda_{m+1/2,2n} - (
O_8\bar{C}_8+C_8\bar{O}_8 ) \Lambda_{m+1/2,2n+1} \ . \label{e3}
\ea
Here the $\Lambda$'s are toroidal lattice sums, while the
orientifold operation is based on $\Omega'=\Omega \Pi_1 $, with
$\Pi_1$ the inversion along the circle, corresponding to the
Klein-bottle amplitude
\be \mathcal{K} \ = \ \frac{1}{2} \, \left \{ (V_8-S_8) W_{2n} \
+\ (O_8-C_8) W_{2n+1} \right\} \ , \label{e4} \ee
where the $W$'s are winding sums, and introduces an $O8_{+}$ plane
at $y=0$ and an $\overline{O8}_{+}$ plane at $y=\pi R$. For
consistency, these demand that no net $R-R$ charge be introduced,
a condition met by N pairs of D8-$\overline{D8}$ branes, where the
choice $N=16$ is singled out by the connection with M-theory
\cite{ads1}. A simple extension of the arguments in \cite{bd}
shows that the unoriented closed spectrum described by (\ref{e3})
and (\ref{e4}) precisely interpolates between the type I string in
the $R \rightarrow \infty$ limit and the type 0B orientifold with
the tachyonic orientifold projection of \cite{bs}, to be
contrasted with the non-tachyonic $0'B$ projection of \cite{ob'},
in the $R \rightarrow 0$ limit.

In order to obtain a classical configuration as in (\ref{t24}),
without tachyons in the open sector, one can put the $N$
$\overline{D8}$ branes on top of the $O8_+$ planes and the $N$
$D8$ branes on top of the $\overline{O8}_{+}$ planes. This
configuration differs from the one emphasized in \cite{ads1} and
related to the phenomenon of ``brane supersymmetry'', with the
$D8$ on top of the $O8_+$ and the $\overline{D8}$ on top of the
$\overline{O8}_{+}$, by an overall interchange of the positions of
branes and anti-branes. This has an important physical effect:
whereas in the model of \cite{ads1} both the $NS$-$NS$ tadpoles
and the $R-R$ charges are {\it locally} saturated at the two
endpoints, in this case there is a local unbalance of charges and
tensions that results in an overall attraction between the
endpoints driving the orientifold system toward a vanishing value
for the radius $R$.

The resulting open string amplitudes\footnote{This model was
briefly mentioned in \cite{ads1} and was further analyzed in
\cite{ablau}.}
\ba \mathcal{A} &=& \frac{N^2 +M^2}{2} \ (V_8-S_8) W_{n} + N M
(O_8-C_8) W_{n+1/2}
\ , \nonumber \\
\mathcal{M} &=&- \frac{N+M}{2} \ \hat{V}_8 W_{n} - \frac{N+M}{2}
\hat{S}_8 (-1)^n W_n \  , \label{e5} \ea
describe matter charged with respect to an $ SO(N) \times SO(M)$
gauge group, where $N=M$ on account of the $R-R$ tadpole
conditions, with nine-dimensional massless Majorana fermions in
the symmetric representations ($N(N+1)/2$,1) and (1,$M(M+1)/2$)
and massive fermions in the bi-fundamental representation ($N,M$).
Notice that tachyons appear for small values of $R$. This spectrum
should be contrasted with the one of \cite{ads1} exhibiting
``brane supersymmetry'', where the massless fermions are in
antisymmetric representations. It is a simple exercise to evaluate
tensions and $R-R$ charges at the two ends of the interval:
\ba
&& T_0\ = \ (N-16) T_8 \ , \qquad T_1 \ = \ (N-16) T_8 \ , \nonumber \\
&& q_0\ = \ - \, (N+16) T_8 \ , \qquad q_1 \ = \ (N+16) T_8 \ .
\label{e6} \ea
These translate into corresponding values for the parameters
$\lambda$ and $\omega$ of the classical solution in (\ref{t23}),
that in this case are
\be \lambda = \frac{1}{R} \ , \quad \cos \omega =
\frac{16-N}{16+N} \ . \label{e7} \ee

Hence, in the new coordinate system (\ref{t025}) the orbifold
operation (\ref{t027}) becomes a $2 \pi$ rotation, and can thus be
related to the fermion parity $(-1)^F$, while the orientifold
operation $\Omega'$ combines a world-sheet parity with a rotation.
Notice also that $\lambda \pi R = \pi$, and therefore the $O_+$
and $\overline{O}_+$ planes are actually juxtaposed in the $(Y,Z)$
plane along the real axis $Y=0$, forming somehow a bound state
with vanishing total $R-R$ charge. To be precise, the $O_+$ plane
lies along the half-line $Y=0,$ $Z>0,$ while the ${\bar O}_+$
plane lies along the complementary half-line $Y=0,$ $Z<0.$ The end
result is that in the $(Y,Z)$ plane one is describing the $0B$ (or
$IIB/(-1)^F$) string, subject to the orientifold projection
$\Omega' =\Omega \Pi_Y $, where $Y,$ as already stressed, is here
a {\it noncompact} coordinate. By our previous arguments, all this
is somehow equivalent to the type IIB orientifold compactified on
a circle that we started with. In more physical terms, the
attraction between the sets of $D(\overline{D})$ branes and
$O(\overline{O})$ planes at the ends of the interval drives them
to collapse into suitable systems of $D$-branes and $O$-planes
carrying no net $R-R$ charge, that should be captured by the
static solutions of the effective action (\ref{t1b}), and these
suggest a relation to the $0B$ theory. In this respect, a
potentially singular fate of space time opens the way to a
sensible string vacuum. We would like to stress, however, that the
picture supplied by the classical solution (\ref{t24}) is
incomplete, since the origin $Y=Z=0$ is actually the site of a
singularity. Indeed, the resulting $O$-plane system has no global
$R-R$ charge, but has nonetheless a dipole structure: its $Z<0$
portion carries a positive charge, while its $Z>0$ portion carries
a negative charge. While we are not able to provide more stringent
arguments, it is reasonable to expect that the condensation of the
open and closed-string tachyons emerging in the $R \to 0$ limit
can drive a natural redistribution of the dipole charges between
the two sides, with the end result of turning the juxtaposed $O$
and $\overline{O}$ into a charge-free type-$O$ orientifold plane.
If this were the case, not only the resulting geometry of the
bulk, but also the $D/O$ systems, would become those of the $0B$
string.

In order to provide further evidence for this, let us look more
closely at the type 0B orientifold we identified, using the
original ten-dimensional construction of \cite{bs}. The 0B torus
amplitude is \cite{dhsw}
\ba \mathcal{T} &=& (|O_8|^2+ |V_8|^2+|S_8|^2 +|C_8|^2)   \ ,
\label{e8} \ea
while the orientifold operation includes the parity
$\Omega'=\Omega \Pi_Y $, so that the Klein-bottle amplitude
\be \mathcal{K} = \frac{1}{2} \ \left ( O_8 + V_8- S_8 - C_8
\right) \ \label{e9} \ee
introduces an $O8$ plane at $Y=0$, without $R-R$ charge and with a
tension that precisely matches that of the type-IIB
$O8$-$\overline{O8}$ bound state. In general the type-0
orientifold planes, being bound states of IIB orientifold planes,
have in fact twice their tension. In the present case, the parity
$\Pi_Y$ along a noncompact coordinate sends one of the orientifold
planes to infinity, with the net result of halving the total
tension seen in the $(Y,Z)$ plane.

One can also add to this system two different types of
brane-antibrane pairs, and the open-string amplitudes read
\cite{bs}
\ba \mathcal{A} &=& \frac{n_o^2 +n_v^2 + n_s^2 +n_c^2}{2} \ V_8 +
(n_o n_v + n_s n_c) \ O_8 \nonumber \\
&-& (n_s n_v + n_c n_o) \ S_8  - (n_s n_o + n_c n_v) \ C_8
 \ , \nonumber \\
\mathcal{M} &=& - \ \frac{n_v + n_o + n_s + n_c}{2} \ \hat{V}_8  \
, \ \label{e10} \ea
while the corresponding $R-R$ tadpole conditions are
\be n_o \ = \ n_v = \ N \ , \quad n_s \ = \ n_c = \ M \ .
\label{e11} \ee

The gauge group of this type-0 orientifold, $SO(N)^2 \times
SO(M)^2$, becomes remarkably similar to that of the type-II
orientifold we started from, provided only branes of one type are
present, together with the corresponding antibranes, a
configuration determined setting for instance $M=0$. The resulting
spectrum is then purely bosonic, and the precise statement is
that, in the $R \rightarrow 0$ limit, the expected endpoint of the
collapse, the spectrum of the type-II orientifold should match the
purely bosonic spectrum of this type-0 orientifold, as was the
case for their closed sectors. Actually, for the geometry of the
$D/O$ configurations this was not totally evident, and the same is
true for the open spectrum, due to an apparent mismatch in the
fermionic content, but we would like to argue again that a proper
account of tachyon condensation does justice to the equivalence.

The open-string tachyon $T_{ai}$ of the type-II orientifold is
valued in the bi-fundamental, and therefore carries a pair of
indices in the fundamental of the $SO(N) \times SO(N)$ gauge
group. In the $R \rightarrow 0$ limit, {\it all} its Kaluza-Klein
excitations acquire a negative mass squared. These tachyons will
naturally condense, with $< T_{ai}
> = T(y) \ \delta_{ai}$, where $T(y)$ denotes the tachyonic kink
profile, breaking the gauge group to its diagonal $SO(N)$
subgroup, so that, after symmetry breaking and level by level, the
fermions will fall in the representations
\ba
&& C ^{(k+1/2)} ~: \frac{N (N-1)}{2} + \frac{N (N+1)}{2} \ , \nonumber \\
&& S ^{(2k)} ~: \frac{N (N+1)}{2} \ , \qquad   S ^{(2k+1)} ~:
\frac{N (N-1)}{2} \ . \label{e12} \ea
In the $R \rightarrow 0$ limit the appropriate description of
tachyon condensation is in the T-dual picture, and after a
T-duality the interactions within the open sector must respect
Kaluza-Klein number conservation. Therefore the Yukawa
interactions, that before symmetry breaking are of the type
\ba && S_{(ij)}^{(2k)} C_{ja}^{t , (k+1/2)}  T_{ai}^{(k-1/2)} \ ,
\qquad
S_{(ab)}^{(2k)} C_{bi}^{(k+1/2)}  T_{ia}^{t , (k-1/2)} \ , \nonumber \\
&& S_{[ij]}^{(2k+1)} C_{ja}^{t , (k+1/2)}  T_{ai}^{(k+1/2)} \ ,
\qquad S_{[ab]}^{(2k+1)} C_{bi}^{(k+1/2)}  T_{ia}^{t , (k+1/2)} \
, \label{e13} \ea
will give rise to the mass terms $S_{(ij)} C_{(ji)}$ , $S_{[ij]}
C_{[ji]}$. The conclusion is that the final low-lying open spectra
are bosonic on both sides and actually match precisely.

A more direct argument for the equivalence we are proposing would
follow from a natural extension of Sen's description of tachyon
condensation \cite{sen}. As we have already stressed, the $O8$ and
$\overline{O8}$ attract one another and drive the orientifold to a
collapse. In the T-dual picture, the $O9$ and $\overline{O9}$
condense into a non-BPS orientifold plane in one lower dimension,
that in the $R \rightarrow 0$ limit becomes the type-0 orientifold
plane that we have described above. This type of phenomenon can
plausibly be related to the closed-string tachyon non-trivial
profile in this model, in a similar fashion to what happens for
the open-string tachyon kink profile in $D$-$\overline{D}$
systems. At the same time, after T-duality the $D9$ and
$\overline{D9}$ branes decay into non-BPS $D8$ branes via the
appropriate tachyon kink profile. Due to the new $(-1)^F$
operation, these new non-BPS type-II branes match directly the
non-BPS type-0 branes discussed in \cite{bs,dms}, since the
$(-1)^F$ operation removes the unwanted additional fermions. Let
us stress that String Theory can resolve in this fashion the
potential singularity associated to an apparent collapse of space
time: after tachyon condensation, the $O$-$\overline{O}$
attraction can give birth to a well defined type-0 vacuum.

In this example one is confronted with the ideal situation in
which a vacuum redefinition can be analyzed to some extent in
String Theory. In general, however, a string treatment in such
detail is not possible, and it is therefore worthwhile to take a
closer look, on the basis of the intuition gathered from Field
Theory, at how the conventional perturbative string setting can be
adapted to systems in need of vacuum redefinitions, and especially
at what it can teach us about the generic features of the
redefinitions. We intend to return to this issue in a future
publication \cite{bdnps}.

\subsection{Threshold corrections and $NS$-$NS$ tadpoles}

While $NS$-$NS$ tadpoles ask for classical resummations that are
very difficult to perform systematically, it is often possible to
identify physical observables for which resummations are needed
only at higher orders of perturbation theory. This happens
whenever, in the appropriate limit of moduli space (infinite tube
length, in the case of disk tadpoles), massless exchanges cannot
be attached to the sources.

The one-loop finiteness of certain types of quantum corrections in
models with broken supersymmetry and $NS$-$NS$ tadpoles was
actually one of the original motivations for our work. One such
example is provided by scalar (Wilson line) masses in
brane-antibrane or brane supersymmetry breaking models
\footnote{Wilson line masses were explicitly computed at one loop
in \cite{abk} in brane supersymmetry breaking models.}. They can
be simply computed from the vacuum energy turning on Wilson lines
$\Lambda (a_i)$, where $A = (e^{2 \pi i a_1} , e^{-2 \pi i a_1},
\cdots e^{2 \pi i a_{16}} , e^{-2 \pi i a_{16}})$ denotes a
collection of Wilson lines in a Cartan subalgebra of the $D$-brane
gauge group under consideration, so that after T-dualities the
$a_i$ can be related to $D$-brane displacements in the internal
space, and the scalar mass matrix is then proportional to
\be {\cal M}_{ij}^2 = {
\partial^2 \Lambda \over \partial a_i \partial a_j} \ . \label{p1}
\ee
Since the propagation of massless modes in the tree-level
(transverse channel) description of (\ref{p1}) is insensitive to
the $D$-brane positions $a_i$, only massive modes contribute to
the mass matrix ${\cal M}_{ij}^2$. As a result, in this particular
example the UV (IR in the tree-level channel) divergence  would
first manifest itself at genus $3/2$.

The threshold corrections to gauge couplings, defined by
\be {4 \pi^2 \over g^2 (\mu) } ={4 \pi^2 \over g_0^2} + \Delta
(\mu, \Phi_i) \ , \label{p4} \ee
where $g_0$ is the tree-level gauge coupling, are a second
interesting observable of this type. These corrections depend on
the energy scale $\mu$ and also on the moduli fields $\Phi_i$ of
the string model under consideration. We can now show that, in a
large class of string constructions with $NS$-$NS$ tadpoles,
including brane-antibrane pairs and brane supersymmetry breaking
models, the one-loop threshold corrections are UV finite, despite
the presence of tadpoles. We can also show that, at the one-loop
order, the differences of gauge couplings for gauge groups related
by Wilson line deformations $W$, $1/g^2 (W) - 1/g^2 (0)$,
quantities that are of direct relevance for unification purposes,
are UV finite in {\it any} non-supersymmetric string model.

The first example that we would like to discuss is obtained adding
N $D5 - \overline{D5}$ pairs to the $T^4/Z_2$ orbifold of the type
$I$ superstring \cite{z2}. If the additional $N$ $D5$ branes are
placed, together with the original 32, at a given fixed point of
the orbifold while the $N$ $\overline{D5}$ are placed at a
different fixed point, that for simplicity we take to be separated
only along one of the internal directions, the resulting gauge
group is $U(16)_9 \otimes [U(16+N) \otimes U(N)]_5$. The N pairs
generate an $NS$-$NS$ tadpole localized in six dimensions, that
would be expected to introduce UV divergences in one-loop
threshold corrections.

In order to obtain a field-theory interpretation, one can turn
windings into momenta via a pair of $T$-dualities that also
convert $D9$ and $D5$ branes into $D7$ and $D3$. Using the
background field construction of \cite{bf}, the one-loop threshold
corrections for the ${D3}$ gauge couplings are then found to be
\cite{bdnps}
\ba  \Delta &=& - \ \frac{4}{v_3} \, ({\rm Tr}\, Q^2)
\int_0^{\infty} \, dl \,
(P^{(2)}-P_e^{(2)}) \  \nonumber \\
&-& (Tr Q^2) \, {N \over 2^8 \pi^2 v_1 v_2 v_3} \int_0^{\infty} dl
\, {\vartheta_2^4 \over
  \eta^{12}} \, \left({2 \pi^2 \over 3} + {\vartheta_2^{''} \over
  \vartheta_2} - {\vartheta_1^{'''} \over 6 \pi \eta^3}\right) \,
  (-1)^m P^{(2)}
  P^{(4)} \, , \label{bsb11}
\ea
where $Q$ is a gauge generator for the $D3$ gauge group,
$v_1,v_2,v_3$ are the volumes of the three internal tori,
$P^{(2)}$ and $P^{(4)}$ are Kaluza-Klein momentum sums along the
torus where the T-duality was performed and along the other two
tori, respectively, $P_e^{(2)}$ is a corresponding even momentum
sum, and $\eta$ and $\vartheta_i$ are Jacobi functions. The
non-supersymmetric contribution in the second line of
(\ref{bsb11}) is IR and UV finite\footnote{For definiteness, here
IR and UV refer to the open (loop) channel.}. The UV finiteness
can be explained from the supergravity point of view, and we shall
return to it below, while the IR finiteness is guaranteed by the
separation between the $D3$ and the $\overline{D3}$ in the
internal space. In the field theory (large volume) limit the
non-supersymmetric contribution is negligible, while the explicit
evaluation of the first term in (\ref{bsb11}) was done in
\cite{bf} and gives
 \be
 \Delta \ = \ - \ {1 \over 4} \ b^{({\cal N}=2)} \ln
({\sqrt{G} \mu^2 |\eta(U)|^4 {\rm Im} U}) \ , \label{bsb011} \ee
where for a rectangular torus of radii $R_1,R_2$,
$\sqrt{G}=R_1R_2$ and ${\rm Im} U= R_1/R_2$. In (\ref{bsb011}),
$b^{({\cal N}=2)}$ denote beta function coefficients for
Kaluza-Klein excitations in the compact torus where the
T-dualities were performed, that fill ${\cal N}=2$ multiplets. The
first, BPS-like contribution in (\ref{bsb11}), is similar to the
standard ${\cal N}=2$ one in orientifold models \cite{bf}, and is
finite. The non-supersymmetric one originates from the cylinder
and reflects the $D3-\overline{D3}$ interaction between branes and
antibranes located at different orbifold fixed points. This
explains, in particular, the origin of the alternating factor
$(-1)^m$. The remarkable property of (\ref{bsb11}) is that the
threshold corrections are UV finite, despite the presence of the
$NS$-$NS$ tadpole. This can be understood noting that in the $l
\rightarrow \infty$ limit the string amplitudes acquire a
field-theory interpretation in terms of dilaton and graviton
exchanges between $Dp$-branes and $Op$-planes. For parallel
localized sources, the relevant terms in the effective Lagrangian
are
\ba S &=& \frac{1}{2 k^2} \int d^{10} x \sqrt {-G} \biggl \{ R -
\frac{1}{2} (\partial \varphi)^2 - \frac{1}{2 (p+2)~!} e^{(5-p-2)
\varphi /2}
F_{p+2}^2 \biggr \} \nonumber \\
&-& \int_{{\bf y} = {\bf y_i}} d^{p+1} \xi  \biggl \{
\sqrt{-\gamma} \left[ T_p  e^{(p-3) \varphi /4} + e^{(p-7) \varphi
/4} \tr F_{\mu \nu}^2 \right] + q C^{(p+1)} \biggr \} \ ,
\label{p5} \ea
where ${\bf \xi}$ are brane world-volume coordinates, $q = \pm 1$
distinguishes between branes or $O$-planes and antibranes or
$\overline{O}$-planes, $G$ is the 10-dimensional metric, $\gamma$
is the induced metric and $C^{(p+1)}$ denotes a $R$-$R$ form that
couples to the branes. The final result for the corrections to
gauge couplings, obtained using (\ref{p5}) while treating for
simplicity the K-K momenta as a continuum, is proportional to
\be T_p \int \frac{d^{9-p} k}{(2 \pi)^{9-p}} \biggl \{ - T ^{\mu
\nu} \, \langle h_{\mu \nu} h_{\rho \sigma}\rangle \, \eta^{\rho
\sigma} \ + \ \tr F^2 \langle \, \varphi \varphi \rangle\,
\frac{(p-3)(p-7)}{16} \biggr \} \ , \label{p6} \ee
where
\be \langle h_{\mu \nu} h_{\rho \sigma}\rangle \ = \ \frac{1}{k^2}
\left( \eta_{\mu \rho} \eta_{\nu \sigma}+ \eta_{\mu \sigma}
\eta_{\nu \rho} - \frac{1}{4} \eta_{\mu \nu} \eta_{\rho
\sigma}\right)  \ee
is the ten-dimensional graviton propagator in de Donder gauge,
\be \langle \varphi \varphi \rangle \ = \ \frac{1}{k^2} \ee
is the ten-dimensional dilaton propagator, and
\be T_{\mu \nu} = \tr \left( F_{\mu \rho} F_{\nu}^{\rho} -
\frac{1}{4} \eta_{\mu
   \nu} F^2 \right)  \label{p7} \ee
is the vector energy-momentum tensor. Irrespective of the dilaton
tadpole, it can be verified that the dilaton and graviton
exchanges in (\ref{p6}) cancel precisely, source by source, in the
threshold corrections, ensuring that the result is actually
finite. Notice that this type of supergravity argument applies to
a large class of string models, including the non-tachyonic type
$0'B$ orientifold and its orbifolds \cite{ob'}.

A second and similar example to this effect is provided by the
brane supersymmetry breaking model proposed in
\cite{sugimoto,bsb}, that contains 32 $D9$ and 32 $\overline{D5}$
branes, for simplicity at the origin of the internal space,
$O9_{+}$ and 16 $O5_{-}$ orientifold planes, so that supersymmetry
is broken at the string scale on the $\overline{D5}$. The main
difference with respect to the previous example is that the model
is tachyon-free for any position of the antibranes in the compact
space. In order to display the decoupling of the tadpole, let us
take for simplicity all internal dimensions to be transverse to
the antibranes, a setting that can be realized performing a pair
of T-dualities in the model of \cite{bsb}, that would turn the
$\overline{D5}$ branes into $\overline{D3}$ and the $D9$ branes
into $D7$. In this case, the one-loop threshold corrections for
the $\overline{D3}$ gauge couplings are found to be \cite{bdnps}
\ba \Delta &=& - \, \frac{4}{v_3} (Tr Q^2) \int_0^{\infty} \ dl \
(P^{(2)}-P_e^{(2)}) \  \nonumber \\
&-& (Tr Q^2) \ {1 \over 8 \pi^2 v_1 v_2 v_3} \int_0^{\infty} dl \
{\vartheta_2^4 \over
  \eta^{12}} \, \left({2 \pi^2 \over 3} + {\vartheta_2^{''} \over
  \vartheta_2} - {\vartheta_1^{'''} \over 6 \pi \eta^3}\right) \ P_e^{(4)}
  P_e^{(2)} \ . \label{bsb12}
\ea
The second, non-supersymmetric contribution to (\ref{bsb12}),
originates from the M\"obius amplitude, but clearly the
combinations of terms appearing in (\ref{bsb11}) and
(\ref{bsb12}), and actually in any similar computation, are the
same. In this example, the non-supersymmetric contribution in the
second line of (\ref{bsb12}) is IR divergent, reflecting the
presence of massless open-string states in the sector that induces
the breaking of supersymmetry. The explicit evaluation of
(\ref{bsb12}) gives
\be \Delta \ = \ {1 \over 4} \ b^{({\cal N}=0)} \, \ln {M_s^2
\over \mu^2} \ - \ {1 \over 4} \ b^{({\cal N}=2)} \ln ({\sqrt{G_i}
\mu^2 |\eta(U)|^4 {\rm Im} U}) \ , \label{bsb012} \ee
where $M_s$ is the string scale and the non-supersymmetric
contribution was here evaluated in the field theory limit, and
where $b^{({\cal N}=0)}$ denotes the contribution from the
supersymmetry breaking sector, so that the total beta function of
the model is\footnote{The beta function is given by the familiar
formula $-11/3 C_2(G) + 2/3 \sum_R T(R) + 1/3 \sum_r T(r)$, where
$C_2(G)$ denotes the Casimir of the adjoint representation for the
gauge group $G$, and $T(R)$ and $T(r)$ are the Dynkin indices for
Weyl fermions and scalars, with an overall normalization such that
$T=1/2$ for the fundamental representation.}
\be b\ = \ b^{({\cal N}=0)} \ + \ b^{({\cal N}=2)} \ = \ 3\ . \ee

Notice also that, in the limit where the internal volume
transverse to the $\overline{D3}$ is large, the second term in
(\ref{bsb11}) and the term in the second line of (\ref{bsb12})
that is most relevant as $l \to \infty$ become negligibly small
compared to the standard contribution from the supersymmetry
breaking sector. The threshold corrections are then dominated by
the first terms, that are precisely given by the standard
supersymmetric expressions \cite{bf}, where only the would-be BPS
winding states contribute and the string oscillators decouple.
Therefore, at the one-loop level, despite the breaking of
supersymmetry at the string scale on the antibranes, the threshold
corrections are essentially determined by a supersymmetric
contribution. This result confirms the conjecture, made in
\cite{bsb}, that in brane-antibrane pairs and in brane
supersymmetry breaking models threshold corrections in codimension
larger than two are essentially given by supersymmetric
expressions, due to the supersymmetry of the bulk (closed string)
spectrum. Another example of this type is provided by a stack of
$D$ branes at orbifold singularities, where the orbifold action on
the Chan-Paton factors breaks the gauge group to $SU(3) \otimes
SU(2) \otimes U(1)$.

As observed in \cite{ls}, the one-loop threshold corrections are
not UV finite in models with intersecting branes
\cite{intersecting}. The supergravity argument in the previous
Section fails in this case since the sources ($D$-branes and
$O$-planes) present in the compact space are not parallel. Since
the brane couplings to bulk fields depend on their spin, the
cancellations that we have described for the case of parallel
branes/orientifolds cease to occur. However, differences of gauge
couplings for gauge groups related by Wilson line breakings $W$,
$1/g^2 (W) - 1/g^2 (0)$, quantities that are of direct relevance
for unification purposes, are still UV finite at one-loop, as can
be seen from general arguments and will be verified explicitly in
\cite{bdnps}.

\section{Conclusions}

Our analysis of tadpoles in Field Theory shows that the endpoint
of the classical resummation flow is an extremum of the effective
potential, but not necessarily a minimum. In addition, the flow
should occur without touching any inflection points, where the
resummation procedure breaks down. In the $\phi^4$ example,
however, we found the peculiar initial points $\phi_0 = - v/2$,
where $v$ are the extrema of the scalar potential, such that only
the first terms in the tadpole expansions contribute and the flows
terminate at the endpoints $\langle \phi \rangle = v$ on the other
side of the potential barrier. Using the link with Newton's
tangent method, we have also associated to this effect a
geometrical interpretation: this occurs whenever the tangent to
the derivative of the potential, $V'$, at the initial ``wrong
vacuum'', crosses the real axis exactly at an extremum. For the
last example considered in Section 2-b one of these points, for
small values of the ``magnetic field'' $c$, lies well within the
convergence region of tadpole resummations, so that the effect is
perturbative. If this finding were to generalize, one could
conceive carrying out the resummation program to the very end both
in Field Theory and in String Theory starting from neighborhoods
of such points. Simple resummations have been performed explicitly
in toy models with localized branes, that affect the bulk fields
via boundary conditions, and we have presented an explicit example
where a toy model for gravity is taken into account at the first
two lowest orders, with a graviton mass term in the wrong vacuum
that is not of Fierz-Pauli type, so that a ghost is present in the
spectrum, while it is of Fierz-Pauli type in the proper vacuum.
Tadpole resummations in the wrong vacuum suffice even in this case
to recover the correct results for physical observables.

At the string level, we have provided some evidence for an example
of explicit vacuum redefinitions connecting a type-$II$
orientifold with $NS$-$NS$ tadpoles to its correct vacuum, that
can be related to a type-0 orientifold. We then showed that,
despite the presence of $NS$-$NS$ tadpoles, one-loop threshold
corrections are UV finite in a large class of string models with
broken supersymmetry, including brane-antibrane pairs, brane
supersymmetry breaking models and essentially any type-$II$ or
type-0 orientifolds with no closed string tachyons (that would
introduce additional divergences in the threshold corrections)
containing parallel BPS or non-BPS brane/orientifold
configurations.

The UV finiteness of the one-loop results presented in Section 3
would seem to imply that the threshold corrections are insensitive
to large tadpoles. This conclusion, however, is likely to be
incorrect, since the one-loop finiteness was the result of a
delicate cancellation between dilaton and graviton exchanges in
the tree-level closed string amplitude. This cancellation is not
expected to be preserved at higher orders, and taking into account
the proper resummations appears to spoil perturbation theory,
yielding a result that, due to the non-analyticity in the dilaton
mass, is of the same order as the tree-level disk contribution.

Starting with genus $3/2$ amplitudes \cite{bsl}, Dyson
resummations of the propagators are clearly necessary. For
example, the higher-genus contributions to Wilson-line masses,
normalized to the one-loop (annulus and M{\"o}bius) contribution
are of order
\be  \frac{(\frac{i}{-V^{\,\prime\prime}})^2\,(-iV^{\,\prime})\,
(-iV^{\,\prime\prime\prime}_{m00}) \, \frac{i}{-m^2}
\,(-iV^{\,\prime})}{\frac{i}{-m^2}\,(-iV^{\,\prime})} \ = \
\frac{V^{\,\prime\prime\prime}_{m00}}{V^{\,\prime\prime}}\,
\frac{V^{\,\prime}}{V^{\,\prime\prime}} \ , \label{p2} \ee
where we considered for definiteness a higher-order string
amplitude whose field theory limit would contain three closed
string propagators, two massless and one massive and one cubic
disk vertex. In (\ref{p2}), $m^2$ denotes the mass of the lowest
Kaluza-Klein excitations and $V^{\,\prime\prime\prime}_{m00}$
denotes the disk three point function for one massive and two
massless closed string fields. Higher-order resummations in this
particular example would thus be perturbative in the tadpole only
if
\be \left| V^{\,\prime\prime\prime}_{m00} V^{\,\prime} /
(V^{\,\prime\prime})^2 \right| << 1 \ . \ee
This condition is not satisfied by the exponential scalar
potential of String Theory, and therefore formally higher-order
contributions would give rise in this case to ${\cal O}(1)$
corrections to the one-loop result. As argued in Section 2-d,
however, for codimension larger than two tadpoles effectively
decouple from the moduli of the transverse space. If, as in the
examples discussed in Section 3, there are additional
contributions from states propagating in only one or two
transverse dimensions that do not couple to the tadpole to leading
order, these are expected to provide the main contributions to
Wilson-line masses or threshold corrections, that are reliably
computed in string perturbation theory. The dependence on the
moduli corresponding to parallel dimensions, on the other hand,
does not decouple.

Concerning the gauge couplings, in the class of models discussed
in Section 3 the one-loop corrections are UV finite. In any
non-supersymmetric model (for instance in intersecting brane
models \cite{intersecting}), only massive states contribute to
differences of gauge couplings for gauge groups related by
Wilson-line deformations $W$, $\Delta (W) \equiv 1/g^2 (W) - 1/g^2
(0)$, the relevant quantities for unification purposes in
Wilson-line GUT breaking models\footnote{The arguments below are
the string analogs of the one-loop UV-insensitive field
theoretical framework proposed in \cite{hw}.}. Consequently, no
tadpole divergences manifest themselves in these quantities at one
loop. In this case, even at higher genus, closed-string
propagators attached to external gauge field lines contain only
massive states. In particular, the genus $3/2$ contribution, when
compared to the one-loop one, is proportional to
$V^{''}_{0m}/V^{''}$, and is therefore renormalizing the one-loop
correction coming from the second, tadpole dependent term in
(\ref{bsb11}). By the previous arguments, for codimension larger
than two these terms are negligible compared to the first term
given by the standard supersymmetric expressions \cite{bf}, to
which only BPS-like states contribute. In principle, the
contributions of parallel dimensions do not decouple, and
therefore the dependence on their moduli is not determined
reliably by computations in the wrong Minkowski vacuum.

The resummation program that we have initiated in this paper can
be of practical use if the endpoint of the resummation flow is a
local minimum of the theory. As we saw in Section 2-b, this is not
always the case, since depending on the starting point a
resummation may well end at a local maximum of the effective
potential or may even not converge at all. Therefore, an important
constraint for tadpole resummations is to start in an appropriate
region, which in String Theory restricts, in particular, the
values of the string coupling. Finally, if non-renormalization
points like those we encountered in Section 2-b were to exist in a
perturbative region and within the convergence radius, ``wrong''
vacua close to them could lead to significant simplifications and
could make resummations not only conceptually important, but also
of real practical use. A proper extension of these resummation
procedures to String Theory must cope with a number of additional
subtleties that are typical of string amplitudes, and is therefore
left for future work \cite{bdnps}.

\newpage
\appendix
\section{Mixed propagators}
Let us consider the ${\cal D}$-dimensional action
\ba S \ &=& \ \int d^{\mathcal D} x \left[
\frac{1}{2}h^{\rho\sigma}\square
h_{\rho\sigma}-h^{\rho\sigma}\dd_\sigma\dd^\mu h_{\rho\mu}
+h^{\rho\sigma}\dd_{\rho}\dd_{\sigma}h-\frac{1}{2}h\square
h+\frac{m^2}{2}
\left( h_{\rho\sigma}h^{\rho\sigma}-ah^2\right)\right] \nonumber\\
&+& \int d^{\mathcal D} x\left[\frac{1}{2}\varphi\left(\square
-M^2\right)\varphi+2b\varphi h\right] \ , \ea
obtained expanding up to the quadratic order the tadpole potential
(\ref{gravity2}), that is also the quadratic part of the
linearized Einstein-Hilbert action coupled to a massive scalar
field, aside from the mass term for the graviton, that is taken to
be of general form. To compute the propagators, one must invert
the kinetic operator
\begin{displaymath}
 \left(\begin{array}{cc}
G_{\mu\nu,\rho\sigma}^{-1} &  -2 i b \ \eta_{\mu\nu} \\
-2 i b \ \eta_{\rho\sigma} & i(p^2+M^2) \\
\end{array} \right) \ ,
\end{displaymath}
where
\ba G_{\mu\nu,\rho\sigma}^{-1} &=& \frac{i}{2}
[(\eta_{\mu\rho}\eta_{\nu\sigma}+\eta_{\mu\sigma}\eta_{\nu\rho})(p^2-m^2)
-2\eta_{\mu\nu}\eta_{\rho\sigma}(p^2-am^2) \nonumber\\
&-&\eta_{\mu\rho}p_{\nu}p_{\sigma}-\eta_{\mu\sigma}p_{\nu}p_{\rho}-
\eta_{\nu\rho}p_{\mu}p_{\sigma}-\eta_{\nu\sigma}p_{\mu}p_{\rho}
+2\eta_{\mu\nu}p_{\rho}p_{\sigma}+2\eta_{\rho\sigma}
p_{\mu}p_{\nu}] \ . \ea
The graviton propagator $G_{\mu\nu}$, the dilaton propagator
$G_{\phi\phi}$ and the mixed propagator $G_{\mu\nu,\phi}$ can be
obtained solving the system
\ba G_{\alpha\beta,\mu\nu}G^{-1\mu\nu,\rho\sigma}\ -\ 2 i b
G_{\alpha\beta,\phi}\eta^{\rho\sigma} &=& \frac{1}{2} \
(\delta^{\rho}_{\alpha}\delta^{\sigma}_{\beta} \ +\
\delta^{\sigma}_{\alpha}\delta^{\rho}_{\beta}) \ , \nonumber\\
-2b\eta^{\mu\nu} G_{\alpha\beta,\mu\nu}\ + \ (p^2+M^2)G_{\alpha\beta,\phi}
&=& 0 \ , \nonumber\\
G_{\mu\nu,\phi}G^{-1\mu\nu,\rho\sigma} \ -\ 2 i b\eta^{\rho\sigma}G_{\phi\phi}
&=& 0 \ ,  \nonumber\\
-2bG_{\mu\nu,\phi}\eta^{\mu\nu}+(p^2+M^2)G_{\phi\phi} &=& -\, i \
. \ea
One can readily fix the general form of the propagators,
\ba G_{\alpha\beta,\mu\nu} &=&
A[\eta_{\alpha\mu}\eta_{\beta\nu}+\eta_{\alpha\nu}\eta_{\beta\mu}+B
\ \eta_{\alpha\beta}\eta_{\mu\nu}
+C(p_{\alpha}p_{\mu}\eta_{\beta\nu}+p_{\alpha}p_{\nu}\eta_{\beta\mu}+p_{\beta}p_{\mu}\eta_{\alpha\nu}
+p_{\beta}p_{\nu}\eta_{\alpha\mu}) \nonumber\\
&+&D(p_{\alpha}p_{\beta}\eta_{\mu\nu}+p_{\mu}p_{\nu}\eta_{\alpha\beta})
+Ep_{\alpha}p_{\beta}p_{\mu}p_{\nu}] \ , \ea
\ba
G_{\alpha\beta,\phi} &=& F \ \eta_{\alpha\beta}+G \ \frac{p_{\alpha}p_{\beta}}{m^2} \ , \\
G_{\phi\phi} &=& H \ , \ea
and the solution is finally
\ba A &=& - \ \frac{i}{2(p^2-m^2)} \ , \qquad B \ =\
\frac{2}{{\mathcal D}-2} \
\left[-1+\frac{(p^2+M^2)(1-2a)m^2+8b^2}{\mathcal Q}\right] \ ,
\nonumber\\
C &=& -\frac{1}{m^2} \ , \qquad D \ = \ \frac{2}{\cal
Q}\left[(p^2+M^2)(1-2a)+\frac{8b^2}{m^2}\right]
 \ , \nonumber\\
E &=& \frac{{\mathcal D} -2}{m^2} \ D \ , \qquad
F \ = \ \frac{2 i b}{\mathcal{Q}} \ , \nonumber\\
G &=&  \frac{2 i b({\mathcal D}-2)}{\mathcal{Q}} \ , \qquad H \ =\
-\, i \  \frac{p^2(1-a)({\mathcal D}-2)+m^2(1-a{\mathcal
D})}{\mathcal{Q}} \ , \ea
where the denominator is
\be \mathcal{Q} \ = \ (p^2+M^2) \ [p^2(1-a)({\mathcal
D}-2)+m^2(1-a{\mathcal D})]+\frac{4b^2}{m^2} \ [p^2({\mathcal
D}-2)+{\mathcal D} m^2] \ . \ee \vskip 36pt
\noindent{\bf Acknowledgements}
We are very grateful to M. Berg, who collaborated with us at the
early stages of this research and contributed most of the figures.
We also benefitted from discussions with C. Angelantonj, D.
Anselmi, C. Bachas, C. Becchi, G. Gabadadze, J. Magnen, J. Mourad,
G. Passarino, M. Porrati, Ya.S. Stanev, and especially with I.
Antoniadis. The work of E.D. was supported in part by the EU
contract HPRN-CT-2000-00148, by the CNRS PICS no. 2530 and by the
EU Excellence Grant MEXT-CT-2003-509661, while the work of M.N.,
G.P. and A.S. was supported in part by INFN, by the EU contracts
HPRN-CT-2000-00122 and HPRN-CT-2000-00148, by the MIUR-COFIN
contract 2003-023852, by the INTAS contract 03-51-6346 and by the
NATO grant PST.CLG.978785. The authors are grateful to the CERN
Theory Department, to the Scuola Normale Superiore of Pisa, to the
CPhT of the {\'E}cole Polytechnique, to the Physics Department of
the University of Rome ``Tor Vergata'' and to the Aspen Center for
Physics for the hospitality extended to them at several stages of
the present work.
\newpage

\end{document}